\documentclass[11pt,a4paper]{article}

\usepackage[margin=25mm]{geometry}
\usepackage[T1]{fontenc}
\usepackage[utf8]{inputenc}
\usepackage{lmodern}
\usepackage{microtype}
\usepackage{amsmath,amssymb,bm,mathtools}
\usepackage{booktabs,array,tabularx,longtable}
\usepackage{graphicx}
\graphicspath{{figures/}}
\usepackage{caption}
\usepackage{subcaption}
\usepackage{enumitem}
\usepackage{xcolor}
\usepackage{hyperref}
\usepackage{bookmark}
\usepackage{float}
\usepackage{placeins}
\usepackage{fancyhdr}

\hypersetup{
  colorlinks=true,
  linkcolor=black,
  citecolor=black,
  urlcolor=blue,
  pdftitle={Adaptive Quadratic Control of Dynamic Systems},
  pdfauthor={Igor Ladnik}
}

\numberwithin{equation}{section}
\newcommand{\R}{\mathbb{R}}
\newcommand{\T}{^{\mathsf T}}
\newcommand{\diag}{\operatorname{diag}}
\newcommand{\sat}{\operatorname{sat}}
\newcommand{\norm}[1]{\left\lVert #1\right\rVert}

\title{\bfseries Adaptive Nonlinear Control with Online Identification and Receding-Horizon Optimization}
\author{Igor Ladnik\thanks{\href{https://www.linkedin.com/in/igor-ladnik-ab93942}{LinkedIn}, \href{mailto:ladnik@hotmail.com}{email}}}
\date{}

\begin{document}
\thispagestyle{plain}
\maketitle
\vspace{-1.5em}

\begin{abstract}
An adaptive nonlinear optimal-control scheme is developed by combining receding-horizon iLQR, state estimation, online parameter identification, and actuator constraints. AMIGO (Adaptive Model-based Intelligent Guidance and Orchestration) organizes the computation into three Time Phases: identification, planning, and closed-loop control. A supervisory adaptive loop monitors predictive consistency during closed-loop operation and can repeat identification and planning when persistent parameter mismatch is detected. The nonlinear transition is evaluated by the fourth-order Runge--Kutta method (RK4), and model parameters are refined by the Levenberg--Marquardt method (LM). The method is illustrated by a Van der Pol oscillator, a quadcopter, and an autonomous lunar-lander descent.
\end{abstract}


\clearpage
\tableofcontents

\clearpage
\section*{Main abbreviations}
\begin{tabularx}{\textwidth}{@{}>{\bfseries}l X@{}}
AMIGO & Adaptive Model-based Intelligent Guidance and Orchestration.\\
EKF & Extended Kalman Filter.\\
iLQR & iterative Linear Quadratic Regulator.\\
KF & Kalman Filter.\\
LM & Levenberg--Marquardt method.\\
LQR & Linear Quadratic Regulator.\\
MPC & Model Predictive Control.\\
RK4 & fourth-order Runge--Kutta method.\\
TP & Time Phase.\\
\end{tabularx}

\clearpage
\section{Introduction}
Quadratic optimal control provides a compact way to balance trajectory-tracking accuracy against control effort. For a linear finite-horizon model, the resulting policy is obtained by a backward dynamic-programming recursion. For a nonlinear model, iLQR repeatedly linearizes the discrete dynamics along a nominal trajectory and solves a sequence of local quadratic problems. A baseline formulation in the notation used here is given in \cite{ladnik2025}; the classical iLQR construction is discussed in \cite{li2004}.

In an adaptive implementation, the optimizer must operate together with state estimation, parameter identification, actuator constraints, and finite computation time. Model parameters can differ from their nominal values, measurements contain noise, and a newly computed policy must be available before the corresponding control instant. AMIGO specifies the ordering of these operations and combines a long-horizon plan with local receding-horizon correction.

\section{AMIGO and Time Phases}
\begin{figure}[H]
\centering
\includegraphics[width=0.98\textwidth]{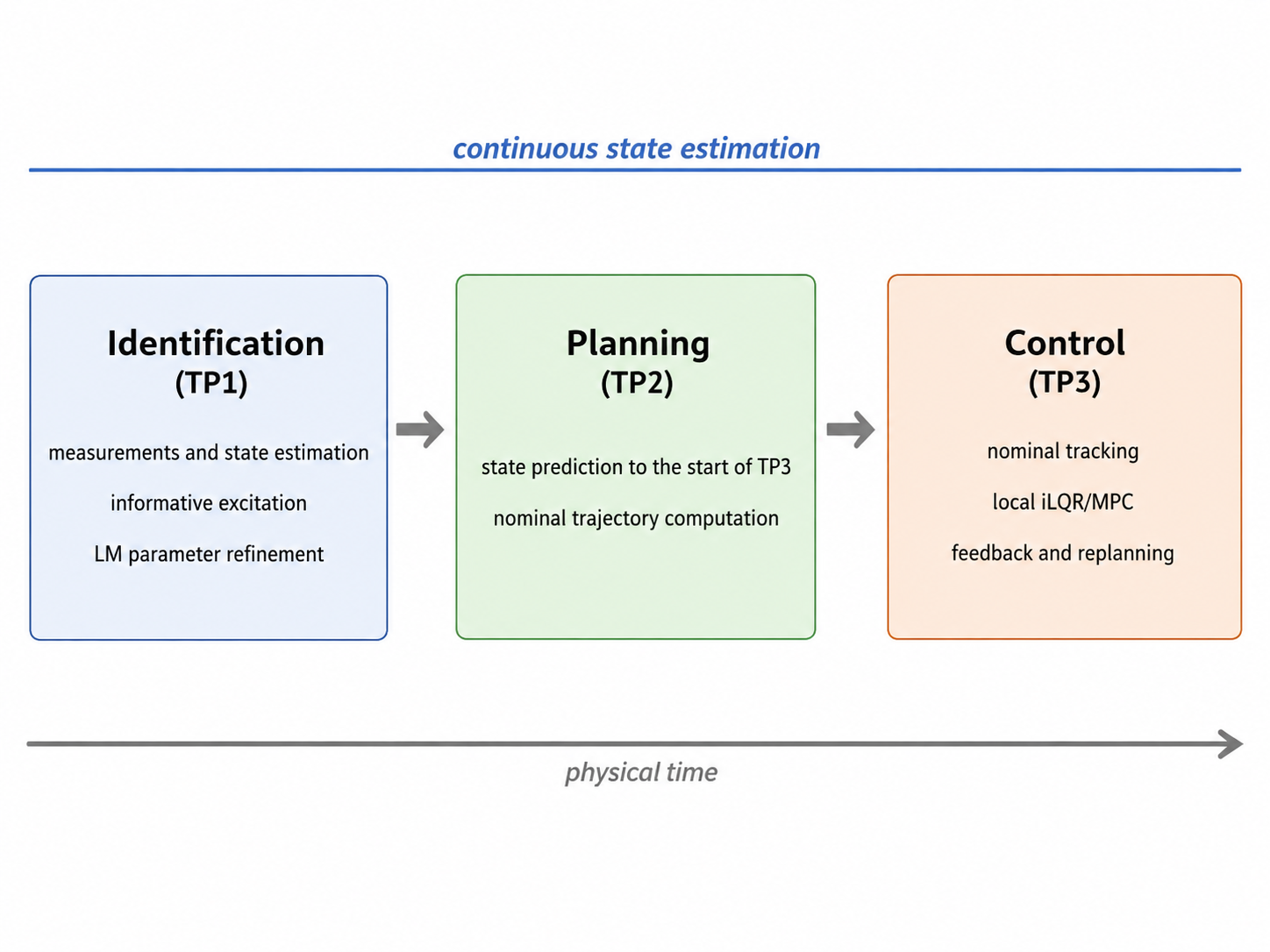}
\caption{AMIGO computation organized into three Time Phases. State estimation continues throughout physical time.}
\label{fig:amigo}
\end{figure}

The abbreviation TP denotes a \emph{Time Phase}. The three phases follow one another in physical time, while state estimation continues throughout the process. Figure~\ref{fig:amigo} summarizes the computation.

\subsection{Identification phase (TP1)}
The plant evolves under a safe baseline control. If the natural motion is not sufficiently informative, a bounded deterministic excitation is added to selected inputs. Measurements are processed by the state estimator, and model parameters are periodically updated from the accumulated history. TP1 terminates when the identification quality is adequate; parameter convergence, residual magnitude, rank, conditioning, and excitation level can all be used in this decision.

\subsection{Planning phase (TP2)}
Physical motion continues during TP2. The identified model predicts the state at the start of TP3. From that predicted state, a long-horizon nominal trajectory is computed. Planning therefore proceeds while the plant is still moving, rather than after the plant reaches a hypothetical stationary handoff point.

\subsection{Control phase (TP3)}
The nominal trajectory becomes the reference for closed-loop control. A shorter local problem with horizon $H$ is periodically solved from the current state estimate. The resulting policy corrects the corresponding segment of the nominal trajectory, and the previous local policy can be reused as a warm start for the next solve.

During TP3, a supervisory mechanism can monitor model consistency and restart identification and planning when persistent mismatch is detected; this adaptive loop is described in Section~\\ref{sec:adaptive-loop}.

\section{Problem formulation and discretization}
Consider the autonomous continuous nonlinear model
\begin{equation}
  \dot{\bm{x}}(t)=\bm{f}\!\left(\bm{x}(t),\bm{u}(t),\bm{\theta}\right),
  \qquad \bm{x}\in\R^n,\quad \bm{u}\in\R^m,
\label{eq:continuous}
\end{equation}
where $\bm{\theta}\in\R^p$ contains unknown or imperfectly known parameters. One integration step of duration $h$ defines the discrete transition
\begin{equation}
  \bm{x}_{k+1}=\mathcal{F}_h(\bm{x}_k,\bm{u}_k,\bm{\theta}).
\label{eq:discrete}
\end{equation}

For trajectory tracking, the finite-horizon cost is written as
\begin{equation}
J=\Phi(\bm{x}_N)+\sum_{k=0}^{N-1}\ell_k(\bm{x}_{k+1},\bm{u}_k,\bm{u}_{k-1}),
\label{eq:cost}
\end{equation}
Throughout the theoretical development, $N$ denotes the number of control intervals. The trajectory therefore contains $N+1$ state samples, $\bm{x}_0,\ldots,\bm{x}_N$, and $N$ controls, $\bm{u}_0,\ldots,\bm{u}_{N-1}$. When the control-rate term is used, $\bm{u}_{-1}$ denotes the known command applied immediately before the optimization horizon.

with
\begin{align}
\ell_k={}&(\bm{x}_{k+1}-\bm{x}_{k+1}^{\star})\T Q_k(\bm{x}_{k+1}-\bm{x}_{k+1}^{\star})
 +(\bm{u}_k-\bm{u}_k^{\star})\T R_k(\bm{u}_k-\bm{u}_k^{\star}) \notag\\
& +(\bm{u}_k-\bm{u}_{k-1})\T S_k(\bm{u}_k-\bm{u}_{k-1}),
\label{eq:stage}
\end{align}
and terminal penalty
\begin{equation}
\Phi(\bm{x}_N)=(\bm{x}_N-\bm{x}_N^{\star})\T Q_f(\bm{x}_N-\bm{x}_N^{\star}).
\label{eq:terminal}
\end{equation}
The weighting matrices satisfy $Q_k\succeq0$, $S_k\succeq0$, $Q_f\succeq0$, and normally $R_k\succ0$. With the convention in \eqref{eq:stage}, the final state $\bm{x}_N$ already appears in the $k=N-1$ stage cost. Thus $Q_f$ in \eqref{eq:terminal} is an \emph{additional} terminal penalty, so the final-state quadratic weight is effectively $Q_{N-1}+Q_f$. The actuator constraints are
\begin{equation}
\bm{u}_{\min,k}\le \bm{u}_k\le \bm{u}_{\max,k},\qquad
\Delta\bm{u}_{\min,k}\le \bm{u}_k-\bm{u}_{k-1}\le \Delta\bm{u}_{\max,k}.
\label{eq:constraints}
\end{equation}

\paragraph{RK4.}
RK4 denotes the fourth-order Runge--Kutta method. Over each sampling interval, the right-hand side of \eqref{eq:continuous} is evaluated at several intermediate points, and the next state is formed from their weighted combination. The same discrete map $\mathcal{F}_h$ is used in the nonlinear rollout and in the local linear model. iLQR requires the discrete Jacobians
\begin{equation}
A_k=\frac{\partial \mathcal{F}_h}{\partial \bm{x}},
\qquad
B_k=\frac{\partial \mathcal{F}_h}{\partial \bm{u}},
\label{eq:jacobians}
\end{equation}
which may be evaluated numerically, analytically for the chosen discretization, or by automatic differentiation.

\section{Quadratic backbone and iLQR}
This section follows the reverse-horizon notation of \cite{ladnik2025}. It is useful here because the dependence of the current control on the remaining future trajectory becomes explicit. To keep the basic recursion transparent, the $\Delta\bm{u}$ term in \eqref{eq:stage} is temporarily omitted; Section~\ref{sec:rate} restores it through an augmented state.

For the linear discrete model
\begin{equation}
\bm{x}_{k+1}=A_k\bm{x}_k+B_k\bm{u}_k,
\label{eq:linear}
\end{equation}
let $m_{N-k}(\bm{x}_k)$ denote the minimum cost over the remaining $N-k$ control steps. Bellman's principle gives
\begin{align}
m_{N-k}(\bm{x}_k)=\min_{\bm{u}_k}\Big[&
(\bm{x}_{k+1}-\bm{x}_{k+1}^{\star})\T Q_k(\bm{x}_{k+1}-\bm{x}_{k+1}^{\star}) \notag\\
&+(\bm{u}_k-\bm{u}_k^{\star})\T R_k(\bm{u}_k-\bm{u}_k^{\star})
+m_{N-(k+1)}(\bm{x}_{k+1})\Big].
\label{eq:bellman}
\end{align}
Assume that the remaining optimal cost can be written as
\begin{equation}
m_{N-k}(\bm{x}_k)
=\bm{x}_k\T P_{N-k}\bm{x}_k
-2\bm{v}_{N-k}^{\mathsf T}\bm{x}_k+a_{N-k},
\label{eq:value}
\end{equation}
where $P_{N-k}$ is symmetric, $\bm{v}_{N-k}$ is a vector, and $a_{N-k}$ does not depend on $\bm{x}_k$. For the terminal penalty \eqref{eq:terminal}, the recursion starts from
\begin{equation}
P_0=Q_f,
\qquad
\bm{v}_0=Q_f\bm{x}_N^{\star}.
\label{eq:terminalPv}
\end{equation}

These terminal conditions follow directly from the explicit terminal penalty \eqref{eq:terminal}. The corresponding constant term is $a_0=(\bm{x}_N^{\star})\T Q_f\bm{x}_N^{\star}$, but it can be omitted because it does not affect the minimizing control. If no separate terminal penalty is used, then $Q_f=0$ and therefore $P_0=0$ and $\bm{v}_0=0$, which recovers the initialization implicit in the formulation of \cite{ladnik2025}. If the terminal target is zero but $Q_f\ne0$, then $\bm{v}_0=0$ while $P_0=Q_f$ remains nonzero.

Substitution of \eqref{eq:linear} and the value function for the remaining $N-(k+1)$ steps into \eqref{eq:bellman}, followed by minimization with respect to $\bm{u}_k$, gives
\begin{equation}
\bm{u}_k=\bm{c}_{N-k}-F_{N-k}\bm{x}_k.
\label{eq:control-law}
\end{equation}
Here $\bm{u}_k$ is the input applied to the open-loop plant. The vector $\bm{c}_{N-k}$ is the state-independent feedforward part of the closed-loop controller, while $F_{N-k}$ is the feedback-gain matrix. Thus the optimizer does not replace the plant input by a different variable: it computes the plant input $\bm{u}_k$ as a feedforward term corrected by state feedback.

Define
\begin{align}
C_k &= Q_k+P_{N-(k+1)},
&\bm{d}_k &= Q_k\bm{x}_{k+1}^{\star}+\bm{v}_{N-(k+1)}, \notag\\
W_{N-k} &= B_k\T C_k B_k+R_k, \notag\\
\bm{c}_{N-k} &= W_{N-k}^{-1}\left(B_k\T\bm{d}_k+R_k\bm{u}_k^{\star}\right), \notag\\
F_{N-k} &= W_{N-k}^{-1}B_k\T C_k A_k.
\label{eq:wcf}
\end{align}
The inverse notation states the algebraic solution; numerically, the corresponding linear systems should normally be solved directly.

Let
\begin{equation}
Z_k=A_k-B_kF_{N-k}.
\label{eq:Z}
\end{equation}
The value-function parameters one step earlier in physical time are then
\begin{align}
P_{N-k} &= Z_k\T C_k Z_k+F_{N-k}^{\mathsf T}R_kF_{N-k}, \notag\\
\bm{v}_{N-k} &= Z_k\T\!\left[\bm{d}_k-C_kB_k\bm{c}_{N-k}\right]
+F_{N-k}^{\mathsf T}R_k\left[\bm{c}_{N-k}-\bm{u}_k^{\star}\right].
\label{eq:Pv-recursion}
\end{align}

Equations \eqref{eq:wcf}--\eqref{eq:Pv-recursion} make the need for the \emph{backward pass} explicit. The coefficients $\bm{c}_{N-k}$ and $F_{N-k}$ at step $k$ depend on $P_{N-(k+1)}$ and $\bm{v}_{N-(k+1)}$, which summarize the optimal cost of the future part of the trajectory. Computation therefore begins from the terminal conditions $P_0,\bm{v}_0$, produces the policy for $k=N-1$, then for $k=N-2$, and continues backward to $k=0$. Only after this reverse sweep are the feedback/feedforward laws available for a forward rollout.

For a nonlinear model, iLQR constructs a nominal trajectory $\{\bar{\bm{x}}_k,\bar{\bm{u}}_k\}$ and linearizes the discrete transition along it:
\begin{equation}
\delta\bm{x}_{k+1}\simeq A_k\delta\bm{x}_k+B_k\delta\bm{u}_k,
\qquad
\delta\bm{x}_k=\bm{x}_k-\bar{\bm{x}}_k,
\quad
\delta\bm{u}_k=\bm{u}_k-\bar{\bm{u}}_k.
\label{eq:local-linear}
\end{equation}
The same reverse-horizon recursion is applied to this local quadratic model, giving
\begin{equation}
\delta\bm{u}_k=\bm{c}_{N-k}-F_{N-k}\delta\bm{x}_k.
\label{eq:local-law}
\end{equation}
In this nonlinear iteration, the desired state and control entering the recursion are expressed relative to the current nominal trajectory. Thus the local stage targets are $\bm{x}_{k+1}^{\star}-\bar{\bm{x}}_{k+1}$ and $\bm{u}_k^{\star}-\bar{\bm{u}}_k$, and the terminal correction-coordinate initialization is
\begin{equation}
P_0=Q_f,
\qquad
\bm v_0=Q_f(\bm x_N^\star-\bar{\bm x}_N).
\label{eq:local-terminalPv}
\end{equation}
Consequently, $\bm{c}_{N-k}$ in \eqref{eq:local-law} is a feedforward \emph{correction}, whereas \eqref{eq:control-law} gives the absolute plant input in the linear tracking problem.
If $W_{N-k}$ is poorly conditioned, regularization is introduced as
\begin{equation}
W_{N-k}\leftarrow W_{N-k}+\lambda I,
\qquad \lambda\ge 0.
\label{eq:regularization}
\end{equation}

After a backward pass, the nonlinear dynamics must be rolled forward to test the new policy. A line search tries scaled feedforward corrections
\begin{equation}
\bm{u}_k^{(\alpha)}=\bar{\bm{u}}_k+\alpha\bm{c}_{N-k}
-F_{N-k}\big(\bm{x}_k^{(\alpha)}-\bar{\bm{x}}_k\big),
\qquad 0<\alpha\le 1.
\label{eq:line-search}
\end{equation}
The coefficient $\alpha$ controls how large a step is taken from the current nominal policy toward the policy obtained in the backward pass. The first trial normally uses $\alpha=1$. If the resulting nonlinear rollout does not reduce the total cost, smaller values are tested. An accepted trial becomes the new nominal trajectory. If none of the tested values is acceptable, the regularization is increased and the backward pass is recomputed.

\section{Control constraints and active set}
\subsection{Amplitude constraints}
At one backward-pass step, the control correction is obtained from a box-constrained quadratic problem of the form
\begin{equation}
\min_{\delta\bm{u}}\;
\delta\bm{u}\T W_{N-k}\delta\bm{u}
+2\bm{g}_k\T\delta\bm{u},
\qquad
\bm{l}_k\le\delta\bm{u}\le\bm{r}_k.
\label{eq:boxqp}
\end{equation}
The bounds are the physical actuator limits expressed in correction coordinates. The vector $\bm{g}_k$ is the linear term of the same local quadratic objective; without bounds its minimizer is $\delta\bm{u}=-W_{N-k}^{-1}\bm{g}_k$. It therefore contains the dependence on the current state deviation and on the future-cost vector $\bm{v}_{N-(k+1)}$.

The active-set procedure starts from the unconstrained local solution. A control component that falls outside its admissible interval is fixed at the nearest bound and declared \emph{active}. The other components remain \emph{free} and are recomputed together. This recomputation matters because $W_{N-k}$ is generally not diagonal: once one actuator is fixed, the best values of the others can change through the cross-coupling terms.

The solver then checks whether the active bounds are consistent with local optimality. If the local derivative indicates that releasing an active variable can reduce the objective, that variable returns to the free set. Conversely, a free variable that reaches a limit becomes active. The add/release cycle continues until all free variables satisfy their unconstrained optimality equations and all active variables satisfy the proper bound conditions. With a fixed amplitude bound, an active input has no local feedback motion away from that bound, whereas the gains of the remaining free inputs are recomputed. When a bound itself depends on the augmented state, as happens for a control-rate limit, the active input follows that affine bound and therefore retains the corresponding bound-induced feedback term. The constrained value function is consequently piecewise quadratic, with each local quadratic piece associated with an active set. This is a compact way to impose box limits directly during the backward pass; a related control-limited formulation is given in \cite{tassa2014}.

\subsection{Control-rate limits}\label{sec:rate}
The penalty
\begin{equation}
(\bm{u}_k-\bm{u}_{k-1})\T S_k(\bm{u}_k-\bm{u}_{k-1})
\label{eq:dupenalty}
\end{equation}
couples adjacent commands and discourages abrupt changes. Because this term contains the previous command, it introduces a state--control cross term and cannot be represented merely by increasing $R_k$.

A convenient exact representation augments the state with the previous control,
\begin{equation}
\bm{z}_k=\begin{bmatrix}\bm{x}_k\\ \bm{u}_{k-1}\end{bmatrix},
\qquad
E=\begin{bmatrix}0&I\end{bmatrix},
\qquad
\bm{u}_{k-1}=E\bm{z}_k.
\label{eq:augstate}
\end{equation}
The corresponding nonlinear augmented transition is
\begin{equation}
\bm{z}_{k+1}=
\begin{bmatrix}
\mathcal{F}_h(\bm{x}_k,\bm{u}_k,\bm{\theta})\\
\bm{u}_k
\end{bmatrix}.
\label{eq:augtransition}
\end{equation}
Around a nominal trajectory $\{\bar{\bm{x}}_k,\bar{\bm{u}}_k\}$, define
$\delta\bm{z}_k=\bm{z}_k-\bar{\bm{z}}_k$ and
$\delta\bm{u}_k=\bm{u}_k-\bar{\bm{u}}_k$, where
$\bar{\bm{z}}_k=[\bar{\bm{x}}_k^{\mathsf T},\bar{\bm{u}}_{k-1}^{\mathsf T}]^{\mathsf T}$.
The local augmented dynamics are
\begin{equation}
\delta\bm{z}_{k+1}=A_k^a\delta\bm{z}_k+B_k^a\delta\bm{u}_k,
\qquad
A_k^a=\begin{bmatrix}A_k&0\\0&0\end{bmatrix},
\qquad
B_k^a=\begin{bmatrix}B_k\\I\end{bmatrix}.
\label{eq:auglinear}
\end{equation}

The nominal command change
\begin{equation}
\bar{\bm r}_k=\bar{\bm u}_k-\bar{\bm u}_{k-1}
\label{eq:nominalrate}
\end{equation}
must be retained in the local quadratic model. Indeed,
$\bm u_k-\bm u_{k-1}=\bar{\bm r}_k+\delta\bm u_k-E\delta\bm z_k$.
Define the local tracking targets
\begin{equation}
\delta\bm u_k^\star=\bm u_k^\star-\bar{\bm u}_k,
\qquad
\delta\bm z_{k+1}^\star=
\begin{bmatrix}\bm x_{k+1}^\star-\bar{\bm x}_{k+1}\\0\end{bmatrix},
\qquad
\widetilde Q_k=\begin{bmatrix}Q_k&0\\0&0\end{bmatrix}.
\label{eq:augtargets}
\end{equation}
For the augmented correction coordinates, the terminal conditions are
\begin{equation}
P_0^a=\begin{bmatrix}Q_f&0\\0&0\end{bmatrix},
\qquad
\bm v_0^a=
\begin{bmatrix}Q_f(\bm x_N^\star-\bar{\bm x}_N)\\0\end{bmatrix}.
\label{eq:augterminal}
\end{equation}
Thus the previous-control component carries no separate terminal penalty.

For a backward step, let
\begin{equation}
C_k^a=\widetilde Q_k+P_{N-(k+1)}^a,
\qquad
\bm d_k^a=\widetilde Q_k\delta\bm z_{k+1}^\star+\bm v_{N-(k+1)}^a.
\label{eq:augCd}
\end{equation}
Expanding the control-tracking and rate terms and minimizing with respect to $\delta\bm u_k$ gives
\begin{align}
W_{N-k}^a &=(B_k^a)\T C_k^aB_k^a+R_k+S_k, \notag\\
\bm c_{N-k}^a &=(W_{N-k}^a)^{-1}
\left[(B_k^a)\T\bm d_k^a+R_k\delta\bm u_k^\star-S_k\bar{\bm r}_k\right], \notag\\
F_{N-k}^a &=(W_{N-k}^a)^{-1}
\left[(B_k^a)\T C_k^aA_k^a-S_kE\right].
\label{eq:augratecontrol}
\end{align}
The term $-S_k\bar{\bm r}_k$ is essential in incremental iLQR coordinates: it accounts for the command change already present in the nominal trajectory. The feedback term $-S_kE$ expresses the coupling to the previous command. Neither term is recovered by the shortcut $R_k\leftarrow R_k+S_k$.

With
\begin{equation}
Z_k^a=A_k^a-B_k^aF_{N-k}^a,
\label{eq:augZ}
\end{equation}
the augmented value-function recursion is
\begin{align}
P_{N-k}^a={}&(Z_k^a)\T C_k^aZ_k^a
 +(F_{N-k}^a)\T R_kF_{N-k}^a
 +(F_{N-k}^a+E)\T S_k(F_{N-k}^a+E), \notag\\
\bm v_{N-k}^a={}&(Z_k^a)\T\!\left[\bm d_k^a-C_k^aB_k^a\bm c_{N-k}^a\right]
 +(F_{N-k}^a)\T R_k\left(\bm c_{N-k}^a-\delta\bm u_k^\star\right) \notag\\
& +(F_{N-k}^a+E)\T S_k\left(\bm c_{N-k}^a+\bar{\bm r}_k\right).
\label{eq:augPv}
\end{align}
Here $P_{N-k}^a\in\R^{(n+m)\times(n+m)}$ and $\bm v_{N-k}^a\in\R^{n+m}$. These equations preserve the same backward-pass logic as Section~4 while correctly retaining the coupling between successive controls.

Hard rate limits are combined with amplitude limits. For the actually applied previous command, the componentwise admissible interval is
\begin{equation}
\max\!\left(\bm u_{\min,k},\bm u_{k-1}+\Delta\bm u_{\min,k}\right)
\le \bm u_k \le
\min\!\left(\bm u_{\max,k},\bm u_{k-1}+\Delta\bm u_{\max,k}\right).
\label{eq:ratebox}
\end{equation}
In correction coordinates, these bounds are shifted by $\bar{\bm u}_k$. During a backward pass the rate bounds are affine in the previous-control component of the augmented state, so an active-set solver should preserve that dependence rather than treating the active rate bound as a state-independent constant. During the forward rollout, the exact interval in \eqref{eq:ratebox} is evaluated from the previous applied command.

\section{State estimation and online identification}
\subsection{State estimation}
The state estimate is updated throughout TP1, TP2, and TP3. The model first predicts
\begin{equation}
\hat{\bm{x}}_{k+1}^{-}=\mathcal{F}_h(\hat{\bm{x}}_k^{+},\bm{u}_k,\hat{\bm{\theta}}_k),
\label{eq:kfpredict}
\end{equation}
and the next measurement corrects this prediction,
\begin{equation}
\hat{\bm{x}}_{k+1}^{+}=\hat{\bm{x}}_{k+1}^{-}
+K_{k+1}\big(\bm{y}_{k+1}-\hat{\bm{y}}_{k+1}^{-}\big).
\label{eq:kfupdate}
\end{equation}
where $\hat{\bm y}_{k+1}^{-}$ is the predicted measurement obtained from the measurement model. The gain $K_{k+1}$ sets the balance between model prediction and measurement according to their error covariances. For nonlinear dynamics, an EKF uses a local linearization of the discrete transition and, when necessary, of the measurement model; the underlying Kalman-filter principle is given in \cite{kalman1960}.

\subsection{Online identification by LM}
LM denotes the Levenberg--Marquardt method \cite{levenberg1944,marquardt1963}. When corrected full-state estimates are used for identification, a one-step model residual can be defined as
\begin{equation}
\bm{r}_k(\bm{\theta})=\hat{\bm{x}}_{k+1}^{+}
-\mathcal{F}_h(\hat{\bm{x}}_k^{+},\bm{u}_k,\bm{\theta}),
\label{eq:idresidual}
\end{equation}
and the parameter estimate is obtained from a bounded weighted least-squares problem,
\begin{equation}
\hat{\bm{\theta}}=
\arg\min_{\bm{\theta}_{\min}\le\bm{\theta}\le\bm{\theta}_{\max}}
\sum_k \norm{\Omega_k^{1/2}\bm{r}_k(\bm{\theta})}^2.
\label{eq:lm}
\end{equation}
where $\Omega_k\succeq0$ is the residual-weight matrix; the symbol $\Omega_k$ is used here to avoid confusion with the control matrix $W_{N-k}$ in the backward pass. LM uses the local sensitivity of the residual to the parameters and damps the update when the linear approximation is insufficiently reliable. During sequential identification, the data history grows while fit quality is checked on a recent validation interval. Residual magnitude, effective rank, conditioning, parameter stability, and excitation level provide useful acceptance diagnostics. With partial observation, a residual formed directly in measurement space is preferable whenever practical because it avoids treating model-generated unobserved state components as independent measurements.

\section{Receding horizon and computational execution}
\subsection{Local replanning}
During TP3, a global trajectory $\{\bar{\bm{x}}_k^{g},\bar{\bm{u}}_k^{g}\}$ provides the nominal reference. A local iLQR problem with prediction horizon $H$ is periodically solved from the current state estimate. Its backward pass uses the same reverse-horizon logic. After line search accepts a local forward rollout, the accepted open-loop sequence already contains the feedforward update, including the accepted value of $\alpha$. It should therefore not be scaled a second time when the stored policy is applied.

When a control-rate penalty is present, define the estimated augmented state and its local nominal value as
\begin{equation}
\hat{\bm z}_k=\begin{bmatrix}\hat{\bm x}_k\\ \bm u_{k-1}^{\mathrm{applied}}\end{bmatrix},
\qquad
\bar{\bm z}_j^{\ell}=\begin{bmatrix}\bar{\bm x}_j^{\ell}\\ \bar{\bm u}_{j-1}^{\ell}\end{bmatrix}.
\label{eq:mpc-augstate}
\end{equation}
For $j=0$, $\bar{\bm u}_{j-1}^{\ell}$ denotes the known command immediately preceding the local horizon. A schematic applied command is then
\begin{equation}
\bm u_k=\sat\!\left(
\bar{\bm u}_j^{\ell}-F_{H-j}^a(\hat{\bm z}_k-\bar{\bm z}_j^{\ell})
\right).
\label{eq:mpc-control}
\end{equation}
Writing $F_{H-j}^a=[F_{x,H-j}\;F_{u,H-j}]$ makes the two feedback contributions explicit:
\begin{equation}
\bm u_k=\sat\!\left(
\bar{\bm u}_j^{\ell}
-F_{x,H-j}(\hat{\bm x}_k-\bar{\bm x}_j^{\ell})
-F_{u,H-j}(\bm u_{k-1}^{\mathrm{applied}}-\bar{\bm u}_{j-1}^{\ell})
\right).
\label{eq:mpc-control-expanded}
\end{equation}
If no control-rate state augmentation is used, the second feedback term is absent and the formula reduces to ordinary state feedback. Here $\sat(\cdot)$ denotes componentwise projection onto the currently admissible control interval; when rate limits are active, that interval is the intersection of the absolute actuator bounds and the interval reachable from the previous applied command. Consecutive local policies may be blended over a short interval so that a newly computed policy does not create an unnecessary command jump.

\subsection{Optimizer computation budget}
The optimizer computation budget is the maximum wall-clock time allowed for a solve. It is independent of the prediction horizon and of the maximum iteration count. A global planning solve can be assigned seconds, while a local update can be restricted to tens or hundreds of milliseconds. If the budget expires, optimization stops and the best trajectory or feedback policy already accepted is retained.

The budget is therefore a direct limit on optimization delay. An iteration limit and an improvement tolerance are separate termination criteria: a solve may finish because further improvement is small, because the maximum number of iterations is reached, or because the time budget is exhausted. In online control, the result must also become available before it is needed; a policy completed later can only affect later control steps.

\subsection{Cost of Jacobian evaluation}
If $A_k$ and $B_k$ are evaluated by central finite differences, a model with $n$ states and $m$ inputs requires approximately $2(n+m)$ additional discrete transitions per trajectory step solely for Jacobian construction. Over a long horizon, this work can exceed the cost of the matrix algebra in the backward pass.

Correct analytical Jacobians of the \emph{discrete} transition can therefore provide a substantial computational gain by eliminating repeated perturbed rollouts. Automatic differentiation is another option. The actual speedup depends on model complexity and on the integration scheme, but it becomes particularly important for long MPC horizons and frequent replanning. Exact-transition caching, warm starts, and shorter local horizons complement this reduction in Jacobian cost.

\section{Supervisory Adaptive Loop}
\label{sec:adaptive-loop}

\subsection{Prediction-consistency monitoring}
The sequence TP1--TP2--TP3 need not be executed only once. During TP3, agreement between the current model and the observed plant behavior can continue to be monitored. This supervisory mechanism closes an outer adaptive loop around the three Time Phases.

At each TP3 step, a computationally inexpensive one-step prediction error is evaluated:
\begin{equation}
\bm e_k^{\mathrm{pred}}=
\hat{\bm x}_{k+1}^{+}
-\mathcal F_h\!\left(\hat{\bm x}_{k}^{+},\bm u_k,\hat{\bm\theta}\right),
\label{eq:adaptive-prediction-error}
\end{equation}
where $\bm u_k$ is the command actually applied to the plant. In implementations with a state estimator, this check normally requires no additional model propagation: the one-step prediction has already been computed by the KF or EKF. When a measurement model is available, the filter innovation---the difference between the measured and predicted outputs---should also be monitored. The resulting statistics can be updated over a sliding window at negligible cost relative to integration, state estimation, and iLQR optimization.

Prediction consistency is the primary diagnostic quantity. Deviation from the nominal or locally replanned trajectory, growth of the realized cost, and persistent actuator saturation provide additional evidence, but trajectory deviation alone is not sufficient to trigger re-identification. It may instead be caused by a transient disturbance, state-estimation error, actuator limitations, or insufficient corrective authority of the local controller.

To avoid expensive diagnostics at every control instant, monitoring is organized in two levels. The first level performs only the inexpensive stepwise checks described above. A second level is activated only when one or more statistics remain outside their admissible ranges. It examines a recent data window, including persistence and temporal correlation of the errors, excitation quality, and whether a trial parameter update improves prediction on validation data.

\subsection{Return to TP1 and TP2}
Three supervisory states are useful. In the \emph{normal} state, TP3 and lightweight monitoring continue. In the \emph{suspect} state, an additional data window is collected and the extended diagnostics are applied. In the \emph{re-identify} state, loss of parametric adequacy has been sufficiently supported and TP1 is restarted. Persistence requirements and separate detection and recovery thresholds provide hysteresis and prevent switching in response to isolated noisy measurements.

During repeated TP1, the plant remains under a safe admissible policy: the preceding policy, a backup stabilizing controller, or a safe baseline command. Bounded excitation may be added only when permitted by the application. Identification can use a recent moving window or accumulated data with increased weight on recent observations; otherwise a large amount of obsolete data may conceal parameter drift.

The updated model is accepted only after checking prediction error on data not used directly for fitting, effective rank, conditioning, parameter stability, proximity to parameter bounds, and excitation sufficiency. If these checks are satisfactory, TP2 is repeated from the current operating situation. It predicts the future TP3 handoff state and prepares a new nominal trajectory and policy, after which closed-loop TP3 operation resumes.

Thus, ordinary operation remains in TP3 and incurs only a small monitoring overhead. Persistent loss of predictive consistency initiates the sequence TP1--TP2--TP3 again rather than relying indefinitely on local replanning with an obsolete model.

\subsection{Parametric mismatch and structural model inadequacy}
Restarting TP1 is useful when the discrepancy can be explained by changed parameters within the assumed model structure, such as mass, inertia, friction, or actuator efficiency. In this case, re-identification should materially reduce the prediction error on new validation data.

A different situation arises when no admissible parameter vector provides sufficient agreement with the observations. Possible indications of structural model inadequacy include prediction errors that remain large after re-identification, improvement on fitting data that does not persist on validation data, systematic bias or temporal correlation in the residuals, residual dependence on state, input, or operating regime, repeated convergence to parameter bounds, and parameter estimates that change substantially with small shifts of the data window. A model that fits one operating regime but fails systematically in another provides similar evidence.

These symptoms can be caused by unmodeled dynamics, delays, actuator nonlinearities, a changed operating regime, unknown external forcing, or insufficient model order. A short observation interval generally cannot prove structural inadequacy conclusively. The diagnostic outcome should therefore be interpreted as detection of \emph{possible} structural inadequacy rather than automatic identification of its cause.

If re-identification with the same model structure does not restore predictive consistency, TP1 should not be repeated indefinitely. Such repetition may merely fit parameters to the current window without improving out-of-sample prediction. The supervisor should instead declare possible structural inadequacy and enter a predefined safe mode. Depending on the application, this may invoke a conservative backup controller, reduce the admissible operating envelope or control intensity, stop the process, or request revision of the model.

Automatic structural switching is possible when a finite set of previously validated models or operating modes is available. Candidate models should be assessed on the same validation window, and a switch should be permitted only when prediction improves sufficiently while stability and safety requirements remain satisfied. Online construction of an unrestricted new model structure is substantially more difficult than parameter identification and lies outside the present scheme.

\section{Examples}
The three examples illustrate one-parameter identification, multi-parameter stabilization, and a long-horizon landing problem with variable mass.

Video visualizations of the quadcopter and lunar-lander examples are supplied with the arXiv submission as the ancillary files
\texttt{amigo\_quadcopter.mp4} and \texttt{amigo\_lunar\_landing.mp4}, respectively.

\subsection{Van der Pol oscillator}
Consider the controlled oscillator
\begin{equation}
\dot{x}=y,\qquad
\dot{y}=\mu(1-x^2)y-x+F.
\label{eq:vdp}
\end{equation}
The control input is the force $F$. The nominal model starts from $\mu_0=1$, whereas the simulated plant uses $\mu=1.4$.

\begin{table}[H]
\centering
\caption{Main data for the Van der Pol example.}
\begin{tabular}{@{}ll@{}}
\toprule
Quantity & Value \\
\midrule
Step, samples, and intervals & $h=0.1$ s; 101 state samples; $N=100$ control intervals \\
Initial state & $(x_0,y_0)=(1,0)$ \\
Desired state & $(x^\star,y^\star)=(0,0)$ \\
Weights & $\diag Q=(1,1)$, $R=1$ \\
Model parameter & $\mu_0=1$, simulated $\mu=1.4$ \\
Control limit & $-5\le F\le 5$ \\
Time Phases & TP1: 0--2.0 s; TP2: 2.0--2.3 s; TP3: from 2.3 s \\
\bottomrule
\end{tabular}
\end{table}

The identification gives $\hat{\mu}=1.40259$, corresponding to an error of approximately $0.185\%$ relative to the simulated value. At the end of the horizon the state is approximately $(1.56\times10^{-3},-1.34\times10^{-2})$.

\begin{figure}[H]
\centering
\includegraphics[width=\textwidth]{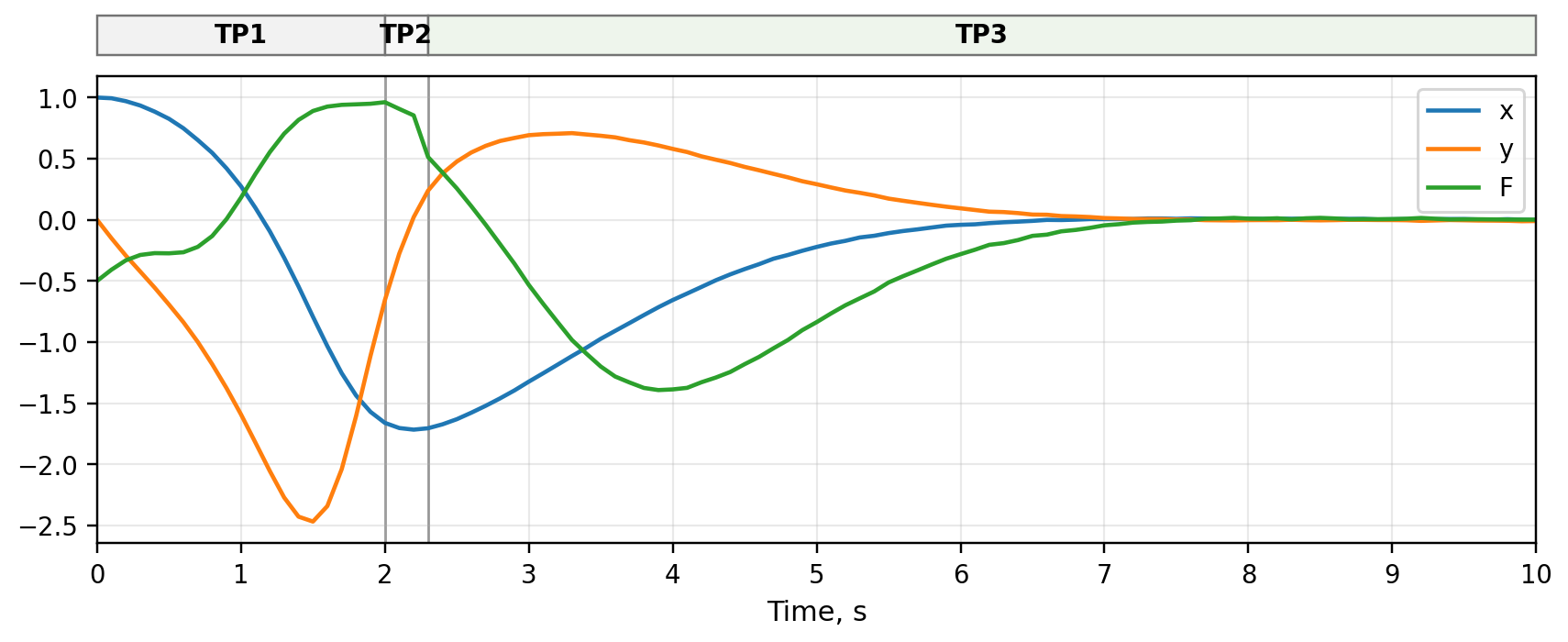}
\caption{Van der Pol oscillator. The phase strip above the plot marks TP1, TP2, and TP3.}
\label{fig:vdp}
\end{figure}

\subsection{Nonlinear quadcopter stabilization}
A 16-state simplified nonlinear quadcopter benchmark is used, with integrators in the actuator channels. The attitude states are the Euler angles $\phi,\theta,\psi$ and three angle-rate states $p,q,r$, propagated in this benchmark as $\dot\phi=p$, $\dot\theta=q$, and $\dot\psi=r$. For finite-angle rigid-body kinematics, the exact transformation between body angular velocity and Euler-angle rates should be used; the present form is a deliberate benchmark approximation.

To keep the rotational dynamics dimensionally consistent, the angular actuator states are interpreted as virtual differential rotor-force channels $f_\phi,f_\theta,f_\psi$ rather than as physical torques. With arm length $l$, the corresponding body moments are $M_\phi=lf_\phi$, $M_\theta=lf_\theta$, and $M_\psi=lf_\psi$ in this simplified model. The rotational equations therefore take the form
\begin{align}
\dot p&=\frac{J_y-J_z}{J_x}qr+\frac{l}{J_x}f_\phi,\notag\\
\dot q&=\frac{J_z-J_x}{J_y}pr+\frac{l}{J_y}f_\theta,\notag\\
\dot r&=\frac{J_x-J_y}{J_z}pq+\frac{l}{J_z}f_\psi.
\label{eq:quadrot}
\end{align}
The four optimization inputs are actuator rates,
\begin{equation}
\dot T=u_T,\qquad \dot f_\phi=u_\phi,\qquad
\dot f_\theta=u_\theta,\qquad \dot f_\psi=u_\psi.
\label{eq:quadact}
\end{equation}
This interpretation preserves the benchmark dynamics while avoiding the dimensional inconsistency that would arise if $f_\phi,f_\theta,f_\psi$ were labeled as torques in equations containing the factor $l/J$.

\begin{table}[H]
\centering
\caption{Selected data for the quadcopter example.}
\small
\begin{tabularx}{\textwidth}{@{}lX@{}}
\toprule
Quantity & Value \\
\midrule
Step, samples, and intervals & $h=0.1$ s; 61 state samples; $N=60$ control intervals \\
Initial angles and rate states & $(\phi,p,\theta,q,\psi,r)=(0.3,0.1,-0.4,0.1,0.2,0.1)$ \\
Initial position and velocity & $(x,v_x,y,v_y,z,v_z)=(0,0.1,0,0.1,0,-0.1)$ \\
Initial actuator states & $T=6.3765$ N, $f_\phi=f_\theta=f_\psi=0$ N \\
Initial parameter estimates & $m=0.65$ kg, $J_x=J_y=0.0075$, $J_z=0.013$ kg m$^2$ \\
Simulated parameters & $m=0.8$ kg, $J_x=0.009$, $J_y=0.0085$, $J_z=0.015$ kg m$^2$ \\
Input limits & $u_T\in[-12,12]$ N/s, $u_\phi,u_\theta,u_\psi\in[-2,2]$ N/s \\
Time Phases & TP1: 0--1.4 s; TP2: 1.4--2.6 s; TP3: from 2.6 s \\
\bottomrule
\end{tabularx}
\end{table}

\begin{table}[H]
\centering
\caption{Identified quadcopter parameters.}
\begin{tabular}{@{}lrrr@{}}
\toprule
Parameter & Simulated value & Estimate & Relative error \\
\midrule
$m$   & 0.800000 & 0.799991 & $-0.001\%$ \\
$J_x$ & 0.009000 & 0.009022 & $0.239\%$ \\
$J_y$ & 0.008500 & 0.008521 & $0.248\%$ \\
$J_z$ & 0.015000 & 0.015061 & $0.403\%$ \\
\bottomrule
\end{tabular}
\end{table}

The global solve requires about 1.10 s with a 1.2 s budget. During TP3, the local problems have a 100 ms budget and reuse the previous policy as a warm start.

\begin{figure}[H]
\centering
\includegraphics[width=0.90\textwidth]{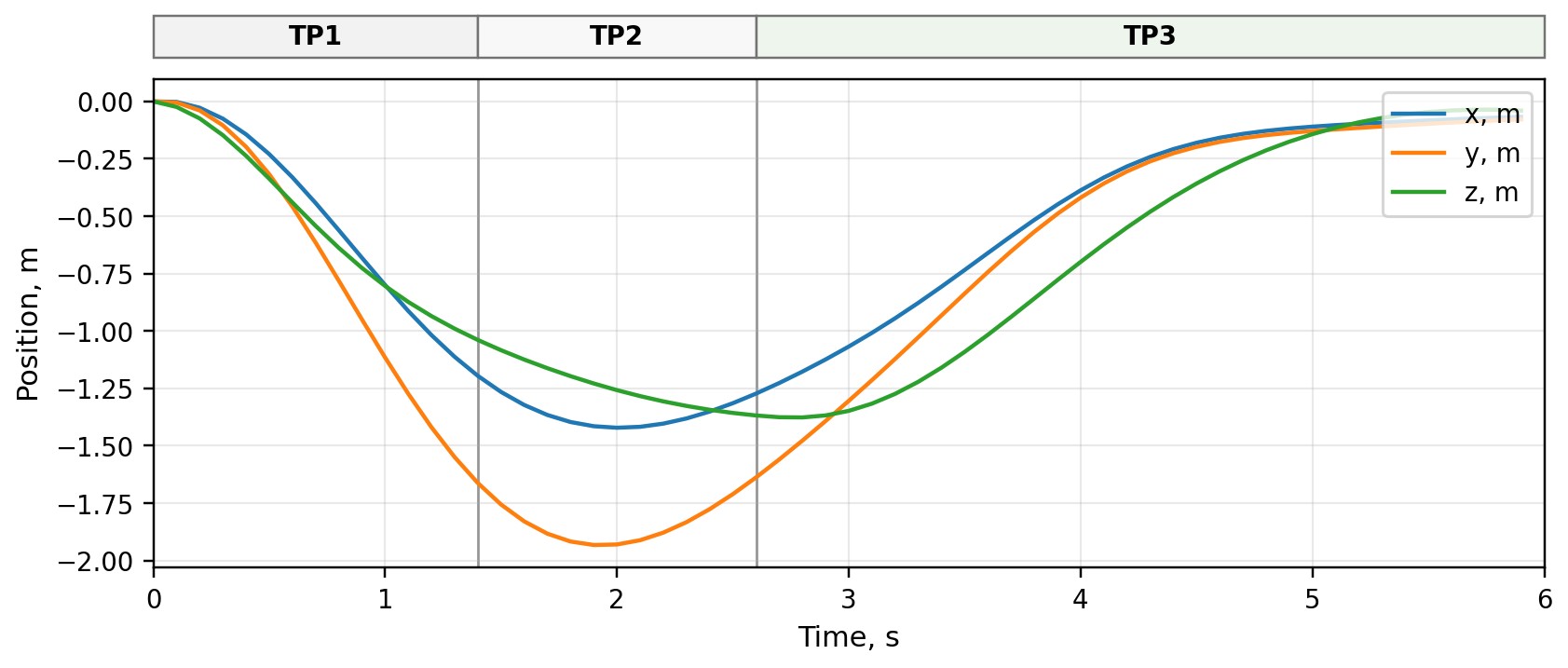}\\[4pt]
\includegraphics[width=0.90\textwidth]{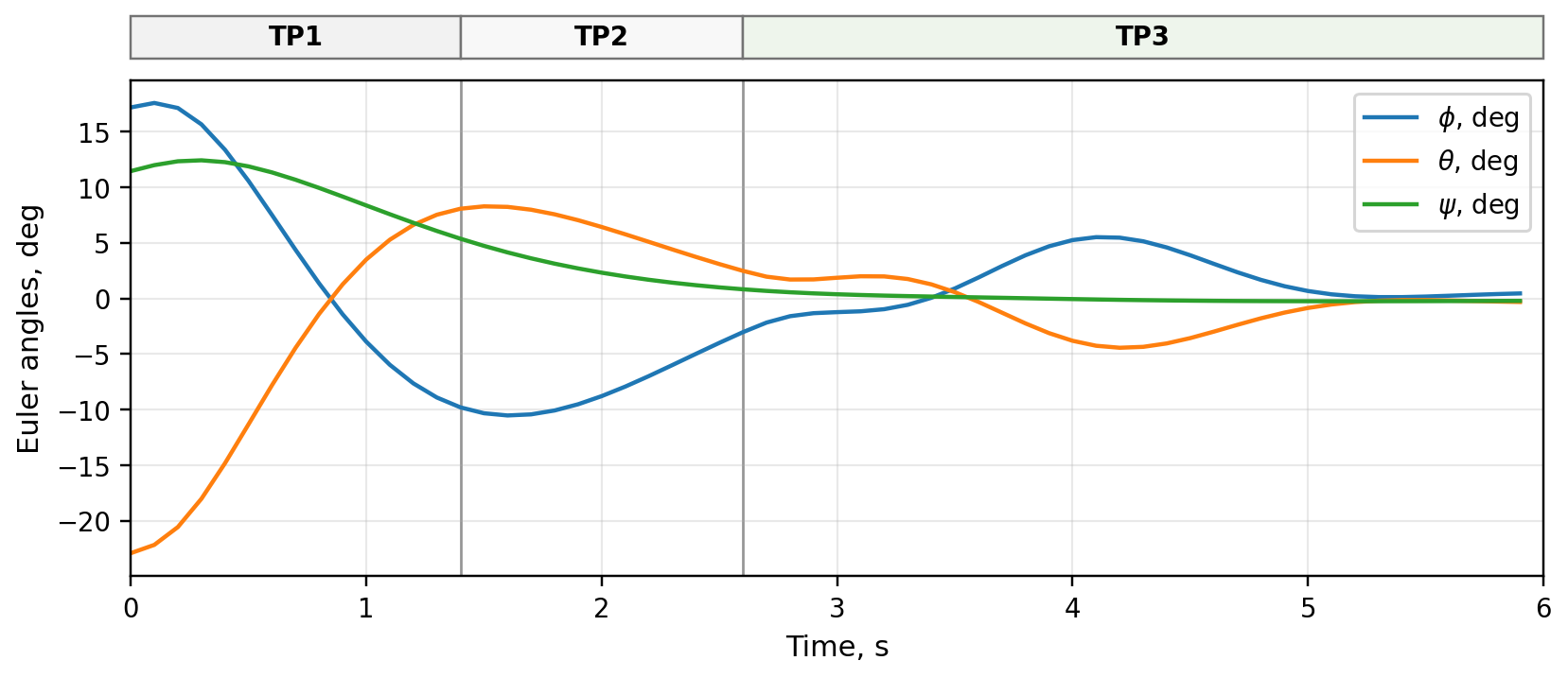}
\caption{Quadcopter position and Euler angles. The phase band is shown above each plot.}
\label{fig:quad-state}
\end{figure}

\begin{figure}[H]
\centering
\includegraphics[width=0.90\textwidth]{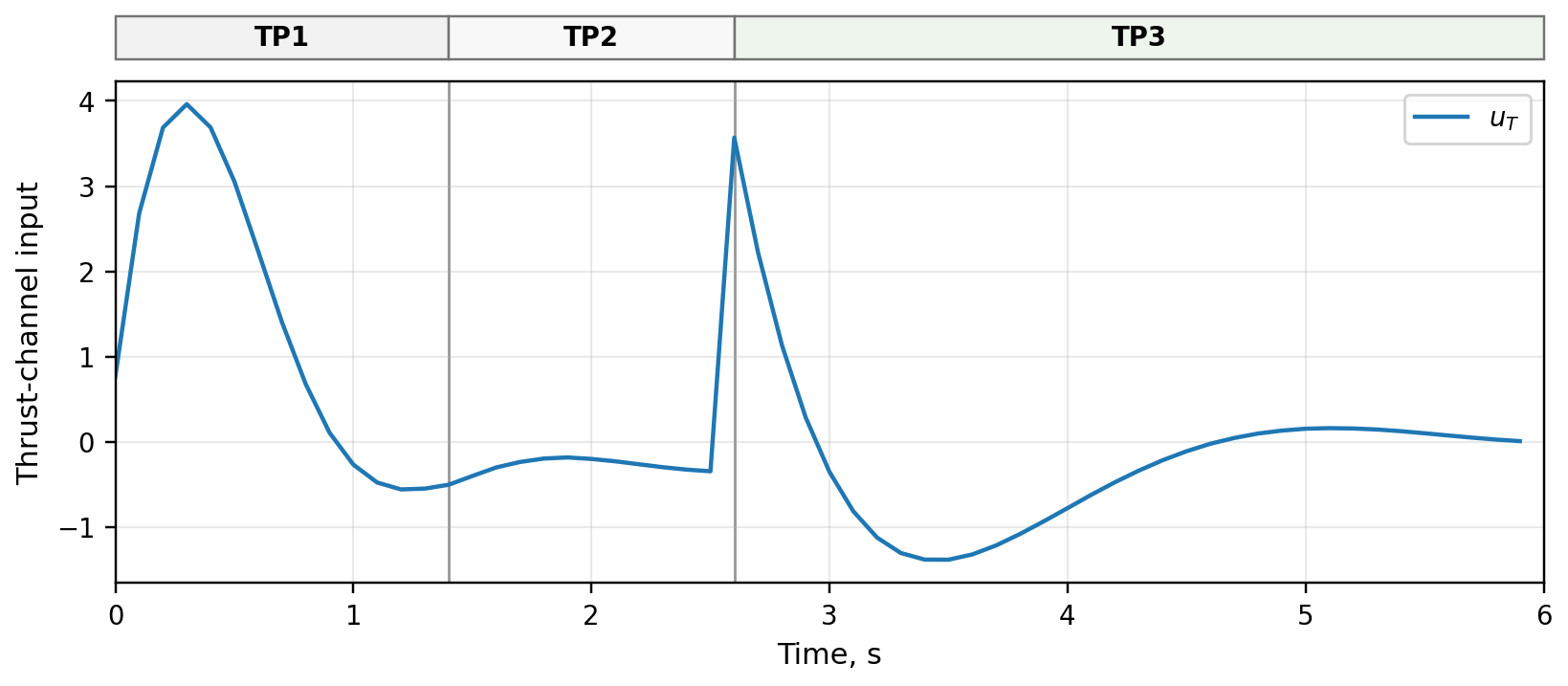}\\[4pt]
\includegraphics[width=0.90\textwidth]{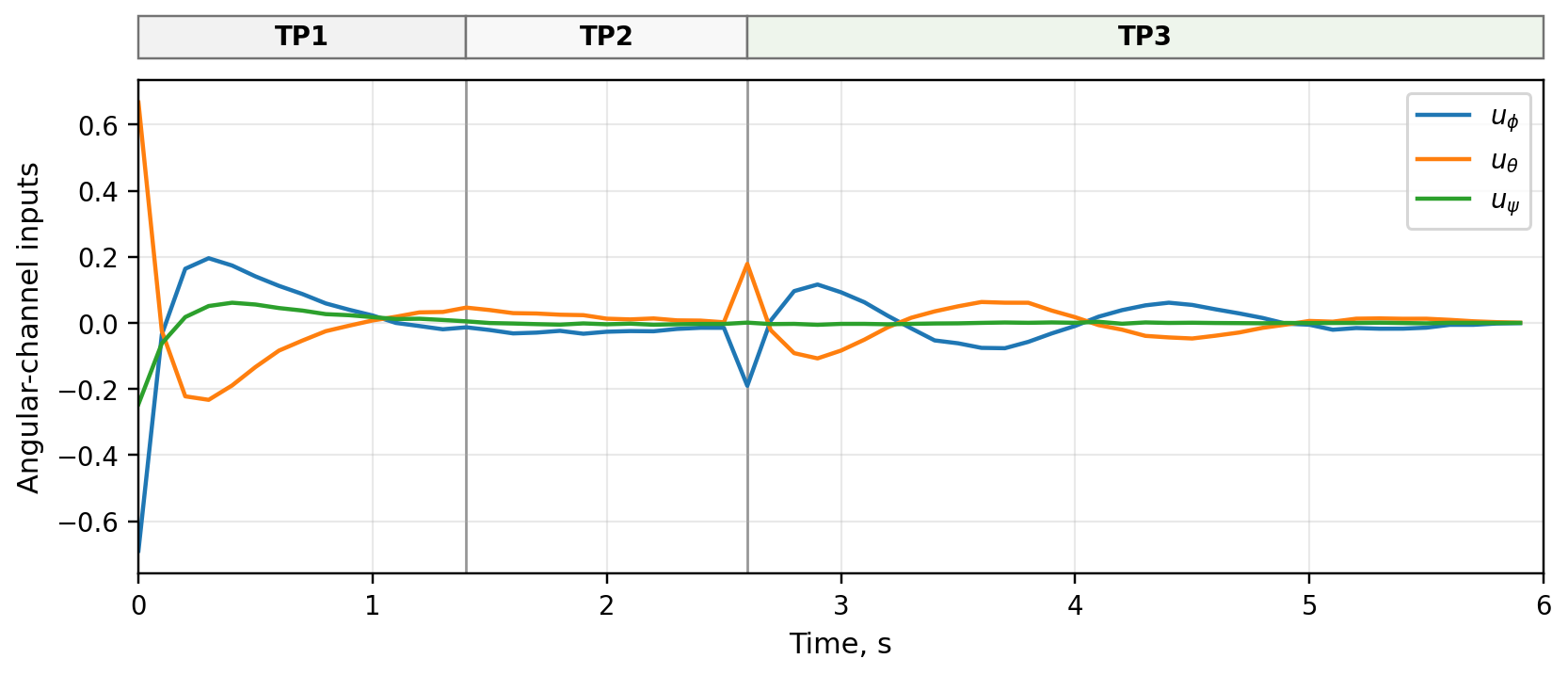}
\caption{Quadcopter actuator-rate inputs: thrust-rate channel $u_T$ and differential-force rate channels $u_\phi,u_\theta,u_\psi$.}
\label{fig:quad-inputs}
\end{figure}

\begin{figure}[H]
\centering
\includegraphics[width=0.78\textwidth]{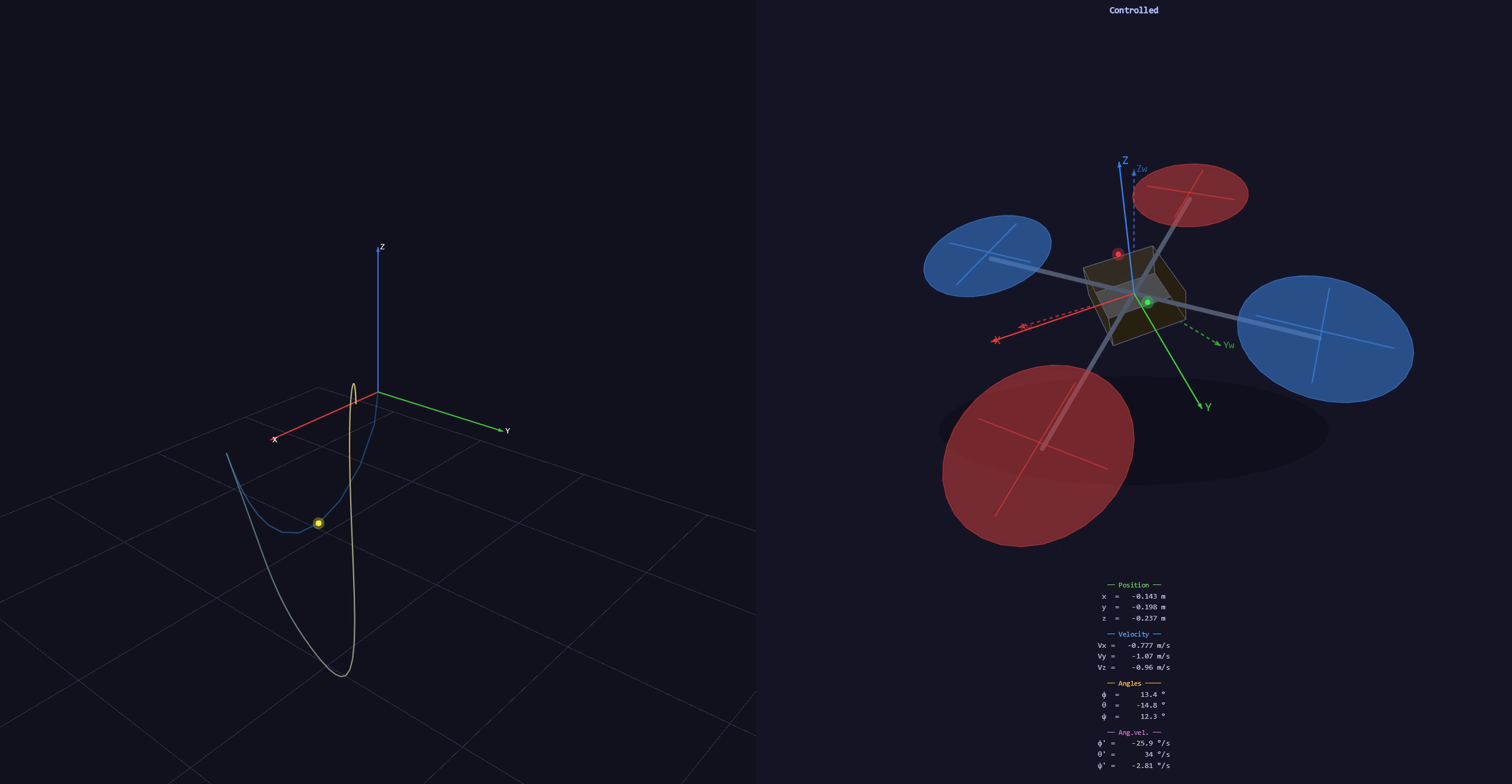}\\[4pt]
\includegraphics[width=0.78\textwidth]{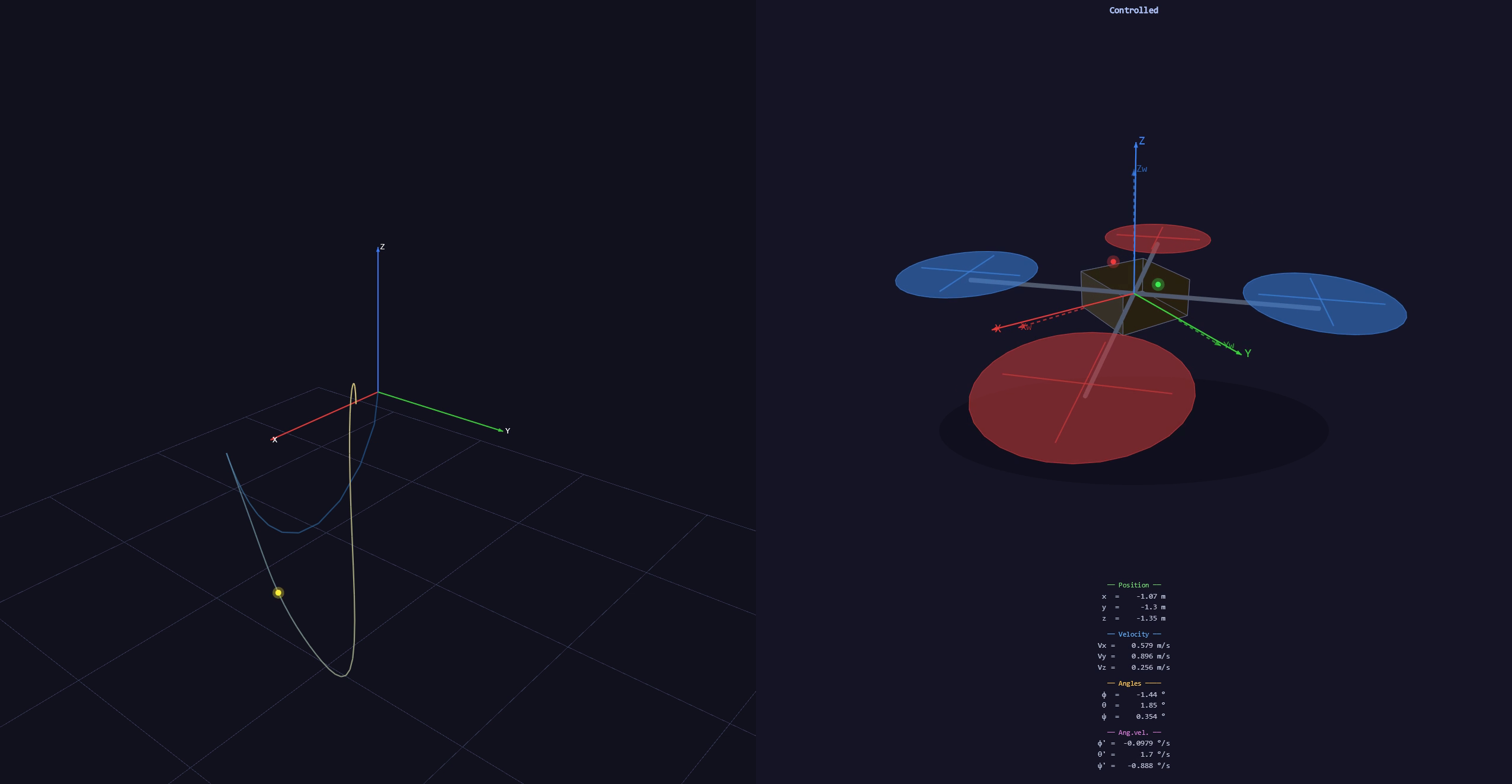}\\[4pt]
\includegraphics[width=0.78\textwidth]{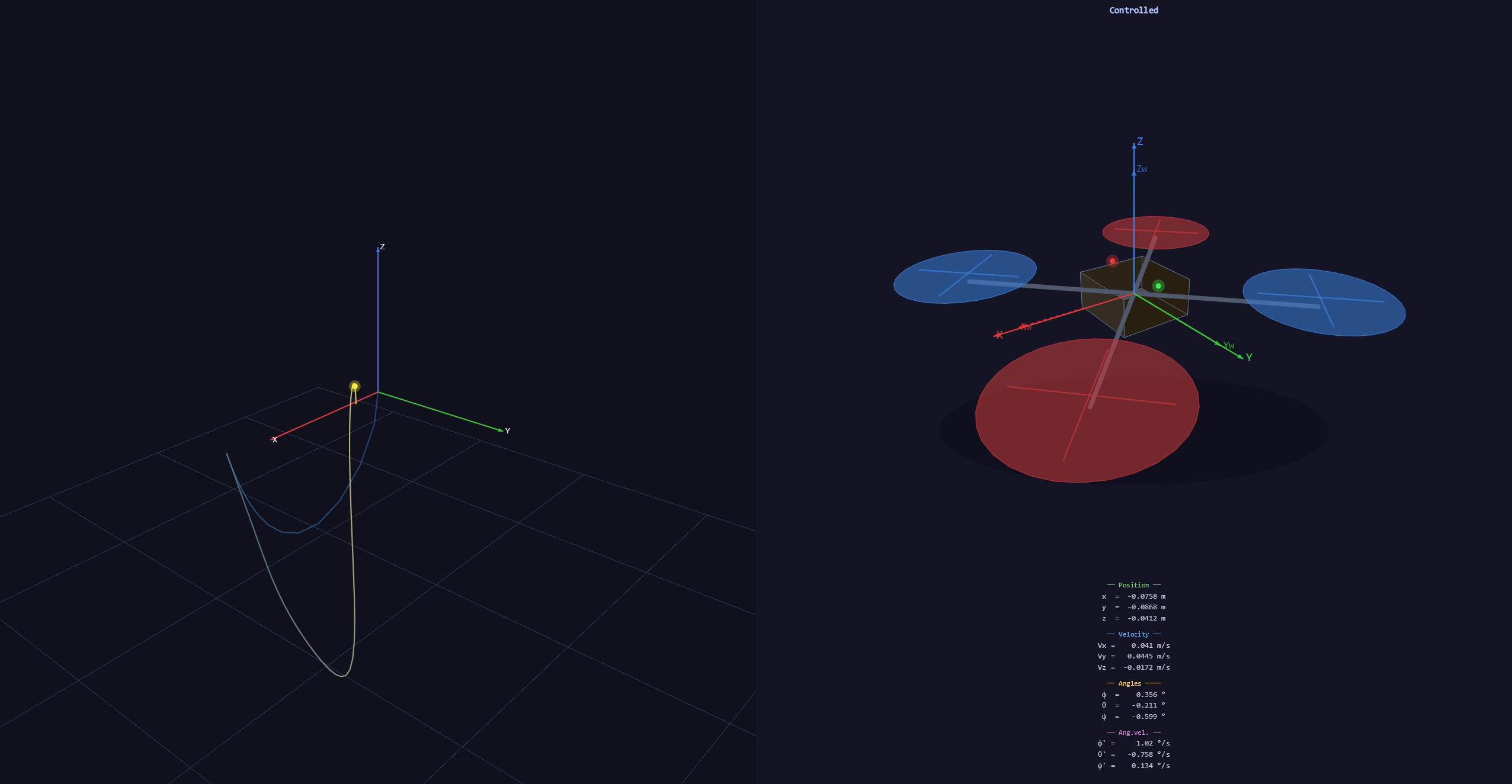}
\caption{Three representative states of the quadcopter motion: beginning, intermediate segment, and end of the horizon.}
\label{fig:quad-frames}
\end{figure}

Over the six-second horizon, the transient is not completely finished at the final sample; the residual vertical speed is approximately $-0.092$ m/s. The obtained trajectory corresponds to the stated finite-horizon weights and constraints.

\subsection{Autonomous lunar-lander descent: a Beresheet-inspired scenario}
Beresheet was an Israeli robotic lunar lander developed by SpaceIL with Israel Aerospace Industries. The Hebrew name \emph{Beresheet} means ``In the Beginning'' and is the opening word of the Book of Genesis. The spacecraft reached lunar orbit and attempted a landing on April 11, 2019; contact was lost shortly before the expected touchdown \cite{nasaBeresheet}.

The scenario uses public mission information to set the physical scale of the main propulsion system. Beresheet used a LEROS~2b engine. The manufacturer specifies a nominal thrust of 420 N and a typical specific impulse of 319.5 s \cite{nammoLeros}; these values are used for $T_{\max}$ and $I_{sp}$. The remaining terminal-descent quantities---attitude-control moments, inertia values, initial mass, and initial state---define the numerical control scenario.

The 14-state vector is
\begin{equation}
\bm{x}=[x,v_x,y,v_y,z,v_z,q_w,q_x,q_y,q_z,\omega_x,\omega_y,\omega_z,m]\T,
\label{eq:lunar-state}
\end{equation}
and the control vector is
\begin{equation}
\bm{u}=[T,\tau_{\mathrm{roll}},\tau_{\mathrm{pitch}},\tau_{\mathrm{yaw}}]\T.
\label{eq:lunar-control}
\end{equation}

\begin{table}[H]
\centering
\caption{Selected data for the lunar-lander descent.}
\small
\begin{tabularx}{\textwidth}{@{}lX@{}}
\toprule
Quantity & Value \\
\midrule
Step, samples, and intervals & $h=0.1$ s; 1707 state samples; $N=1706$ control intervals \\
Initial position & $(x,y,z)=(25,-20,800)$ m \\
Initial velocity & $(v_x,v_y,v_z)=(4,-3,-8)$ m/s \\
Initial quaternion & $(q_w,q_x,q_y,q_z)=(0.90566,0.32498,-0.15231,0.22579)$ \\
Initial angular rates & $(\omega_x,\omega_y,\omega_z)=(0.02618,-0.01745,0.01396)$ rad/s \\
Initial mass & $m_0=177.1$ kg \\
Lunar gravity and main engine & $g=1.62$ m/s$^2$, $T_{\max}=420$ N, $I_{sp}=319.5$ s \\
Torque limits & $\tau_{\mathrm{roll}},\tau_{\mathrm{pitch}}\in[-30,30]$ N m, $\tau_{\mathrm{yaw}}\in[-20,20]$ N m \\
Time Phases & TP1: 0--15 s; TP2: 15--25 s; TP3: from 25 s \\
\bottomrule
\end{tabularx}
\end{table}

During TP1, the efficiency factors of the main-thrust and three attitude-control channels are identified.

\begin{table}[H]
\centering
\caption{Identification of actuator-efficiency factors.}
\begin{tabular}{@{}lrrr@{}}
\toprule
Parameter & Simulated value & Estimate & Error \\
\midrule
$\eta_T$ & 0.940000 & 0.940316 & $0.034\%$ \\
$\eta_{\mathrm{roll}}$ & 1.030000 & 1.029683 & $-0.031\%$ \\
$\eta_{\mathrm{pitch}}$ & 0.900000 & 0.900052 & $0.006\%$ \\
$\eta_{\mathrm{yaw}}$ & 0.970000 & 0.970897 & $0.092\%$ \\
\bottomrule
\end{tabular}
\end{table}

\begin{figure}[H]
\centering
\includegraphics[width=0.78\textwidth]{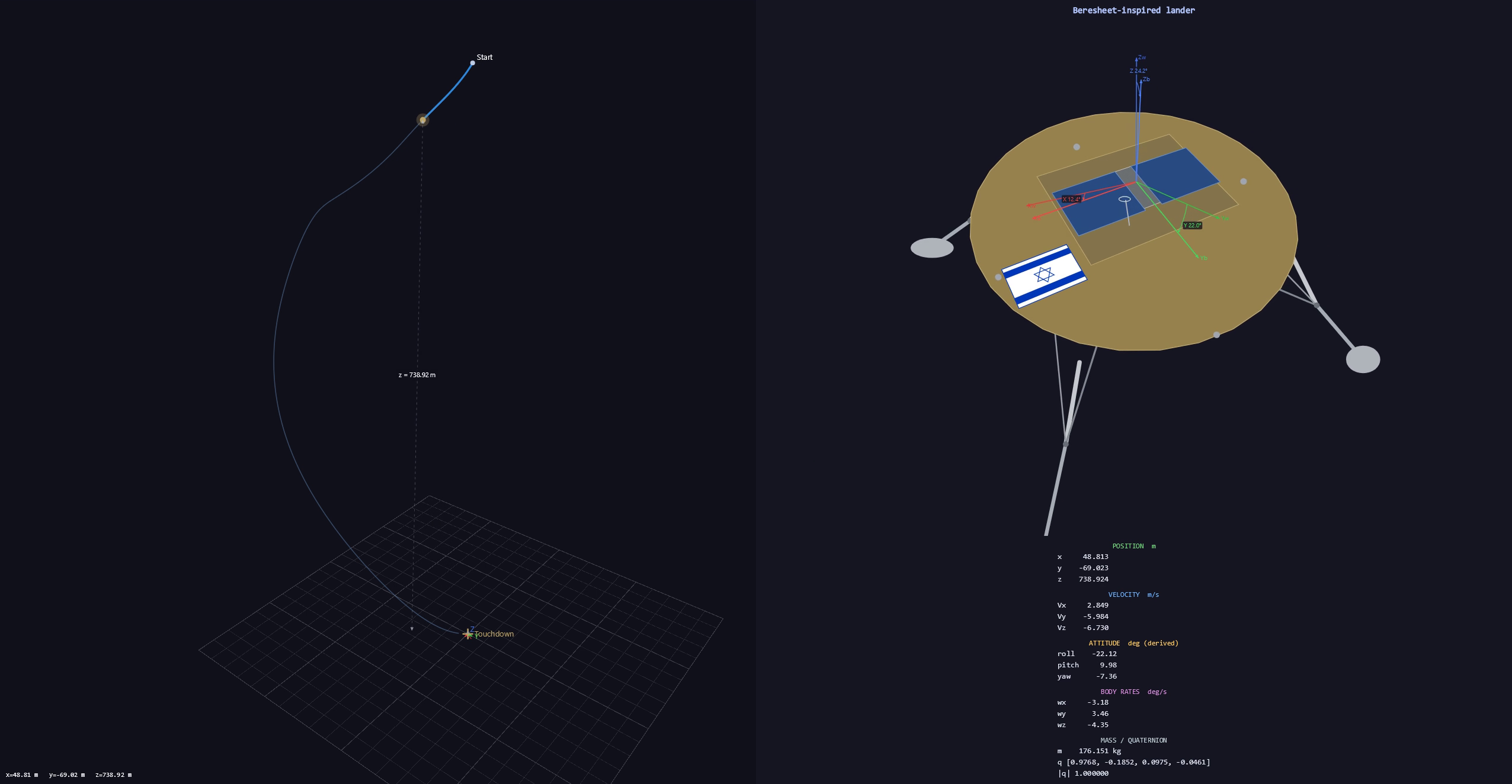}\\[4pt]
\includegraphics[width=0.78\textwidth]{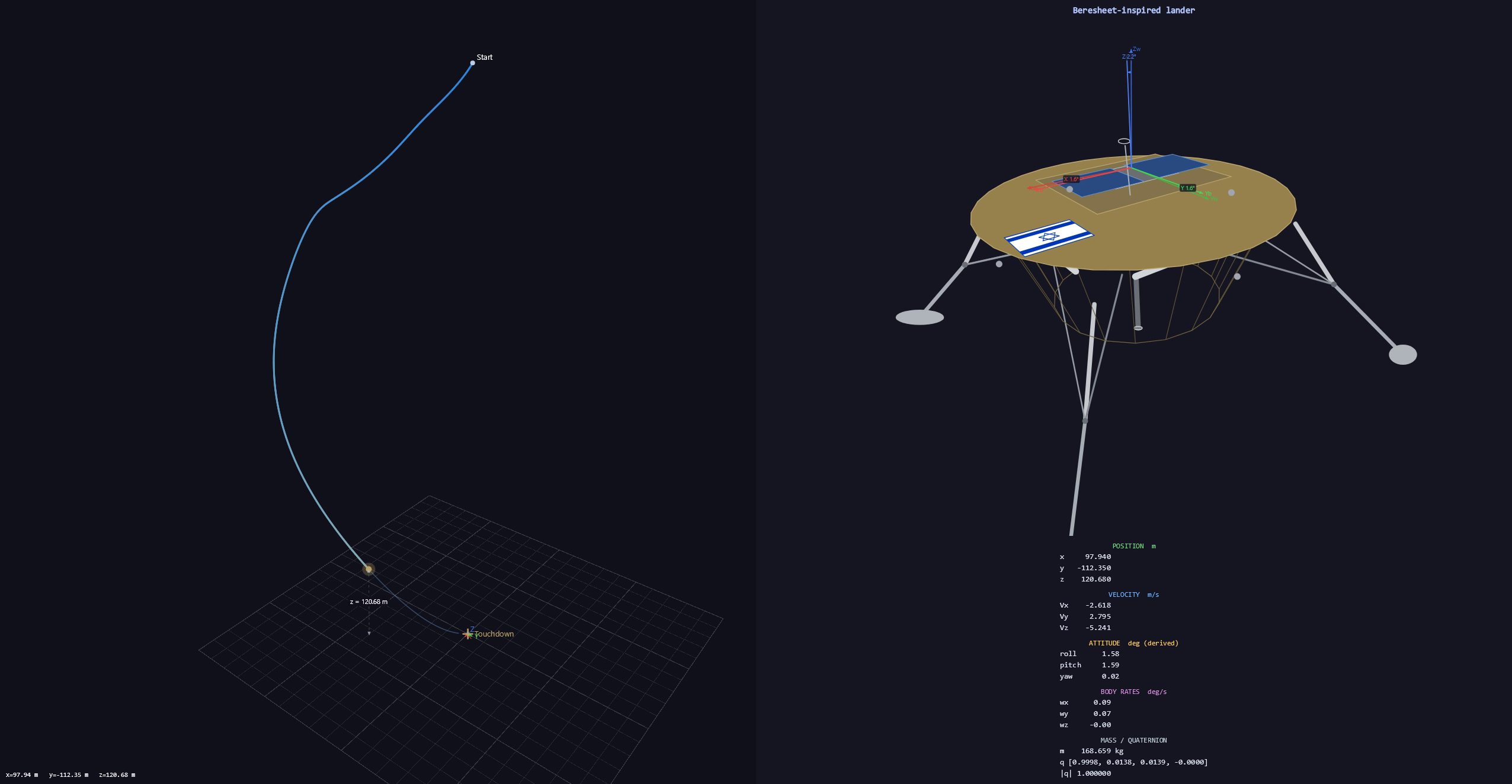}\\[4pt]
\includegraphics[width=0.78\textwidth]{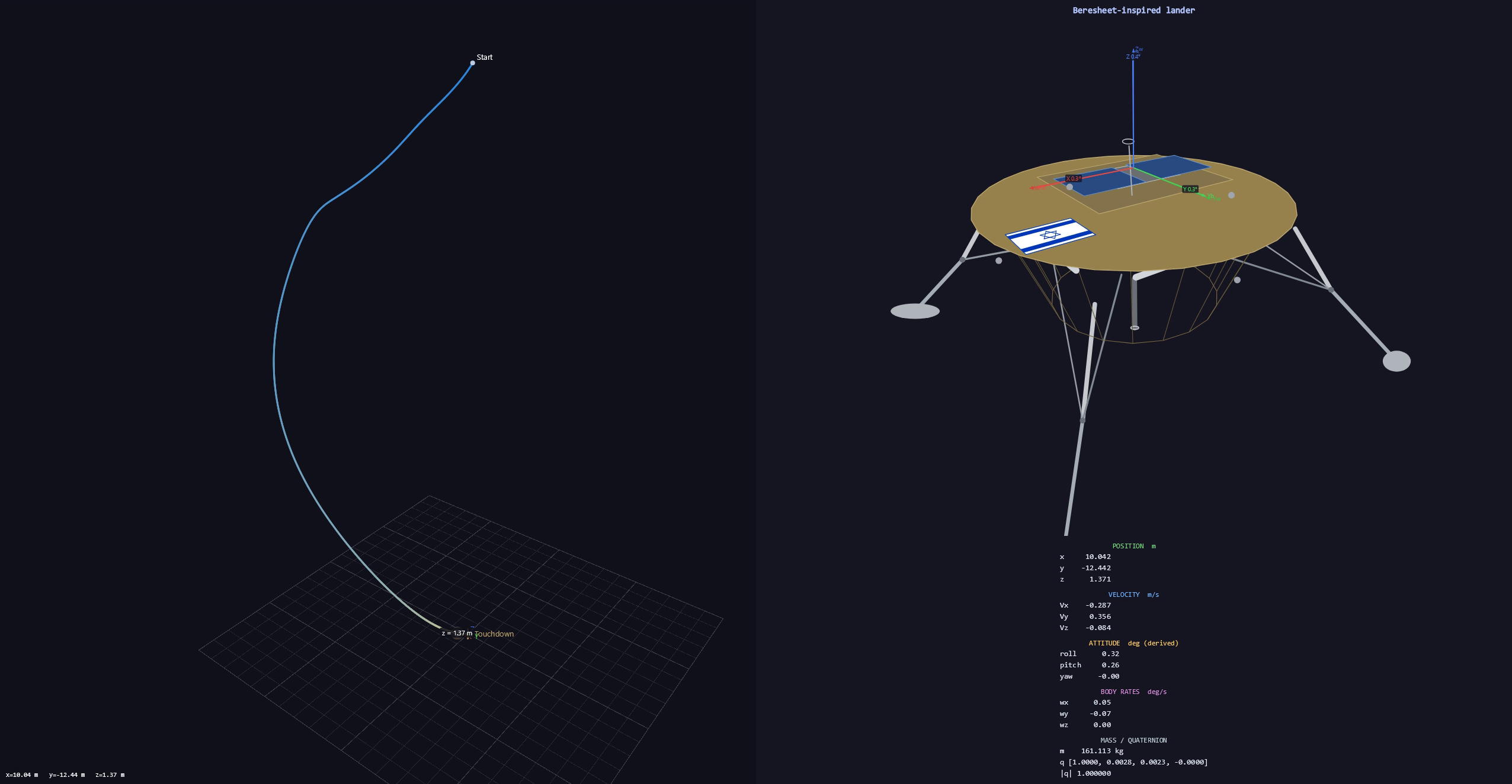}
\caption{Three representative states of the lunar descent: early segment, mid-trajectory, and approach to the surface.}
\label{fig:lunar-frames}
\end{figure}

\begin{figure}[H]
\centering
\includegraphics[width=0.90\textwidth]{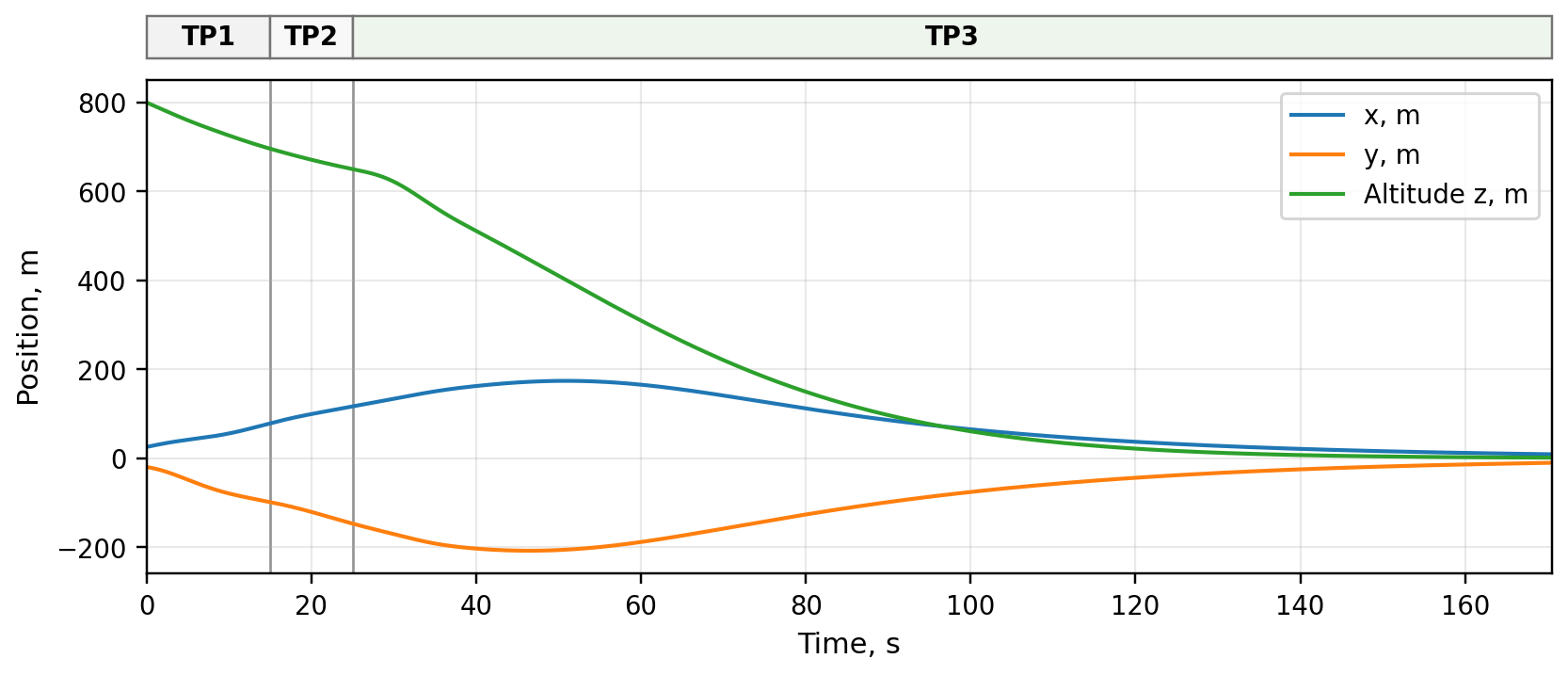}\\[4pt]
\includegraphics[width=0.90\textwidth]{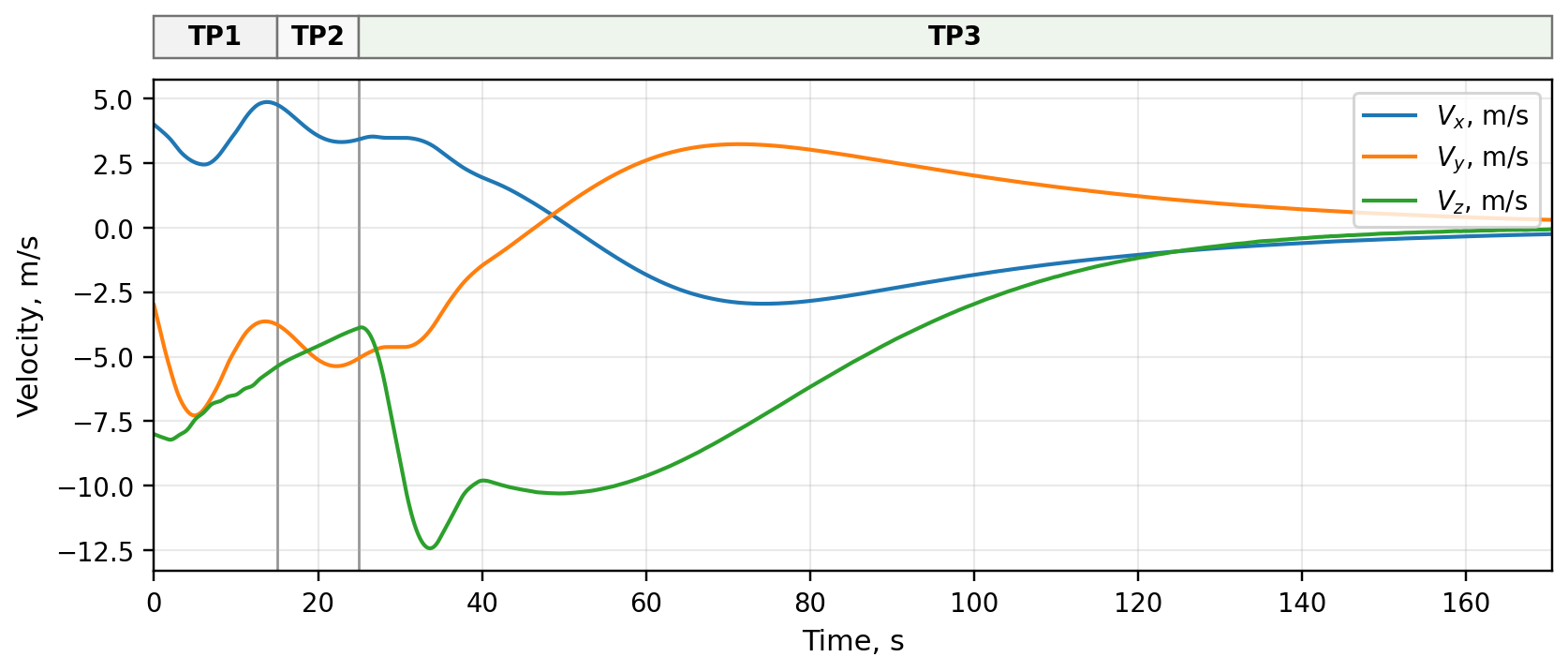}
\caption{Lunar-lander position and translational velocity.}
\label{fig:lunar-posvel}
\end{figure}

The altitude decreases from 800 m to approximately 0.95 m, while the vertical speed reaches approximately $-0.0582$ m/s at the final sample. The final horizontal coordinates are approximately $x=8.54$ m and $y=-10.60$ m, with horizontal velocities about $-0.245$ m/s and $0.305$ m/s, respectively. The example demonstrates a controlled soft approach together with attitude stabilization.

\begin{figure}[H]
\centering
\includegraphics[width=0.90\textwidth]{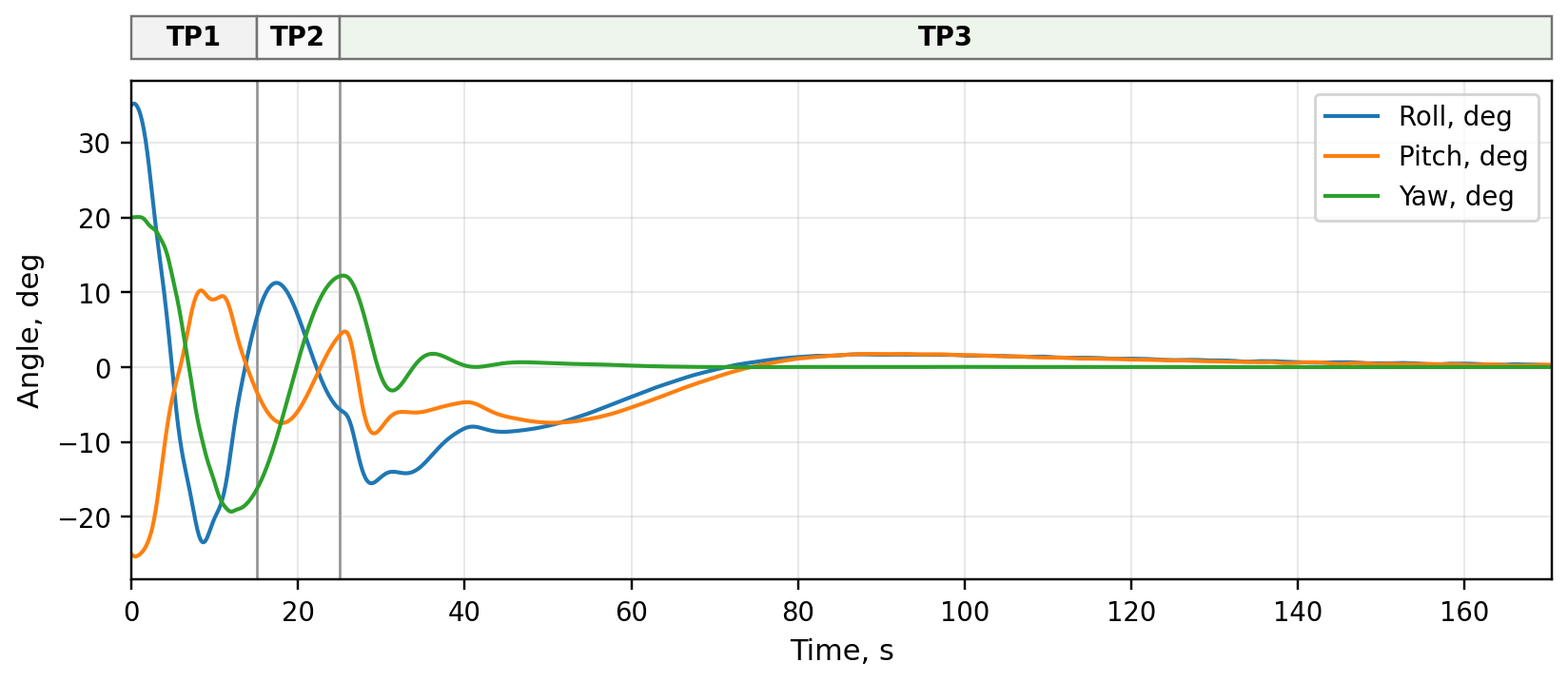}\\[4pt]
\includegraphics[width=0.90\textwidth]{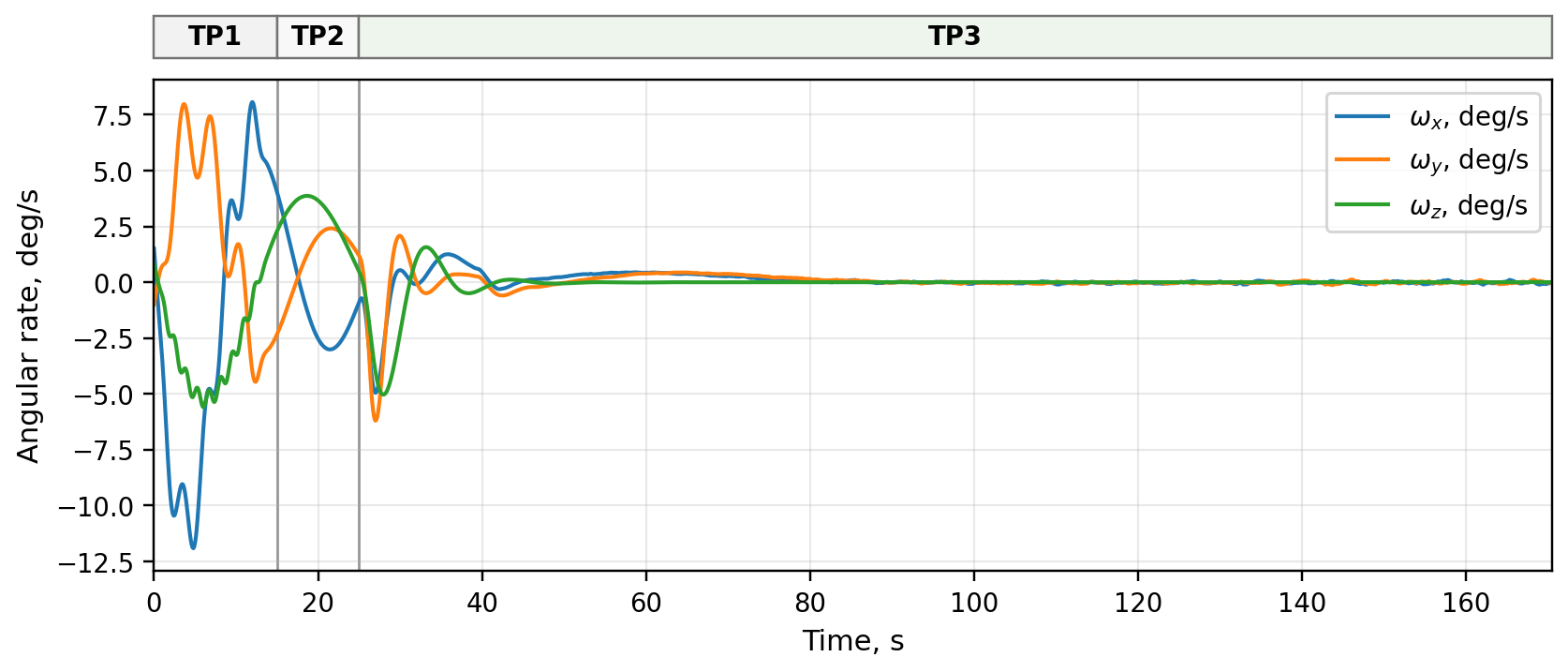}
\caption{Lunar-lander Euler angles and angular rates. The optimized attitude itself is represented by a quaternion; Euler angles are shown for visualization.}
\label{fig:lunar-att}
\end{figure}

The attitude-control torques are initially large because the vehicle must reduce the initial attitude error and angular rates. After the main attitude transient, the angular channels provide smaller corrections that maintain the direction of the main thrust vector.

\begin{figure}[H]
\centering
\includegraphics[width=0.90\textwidth]{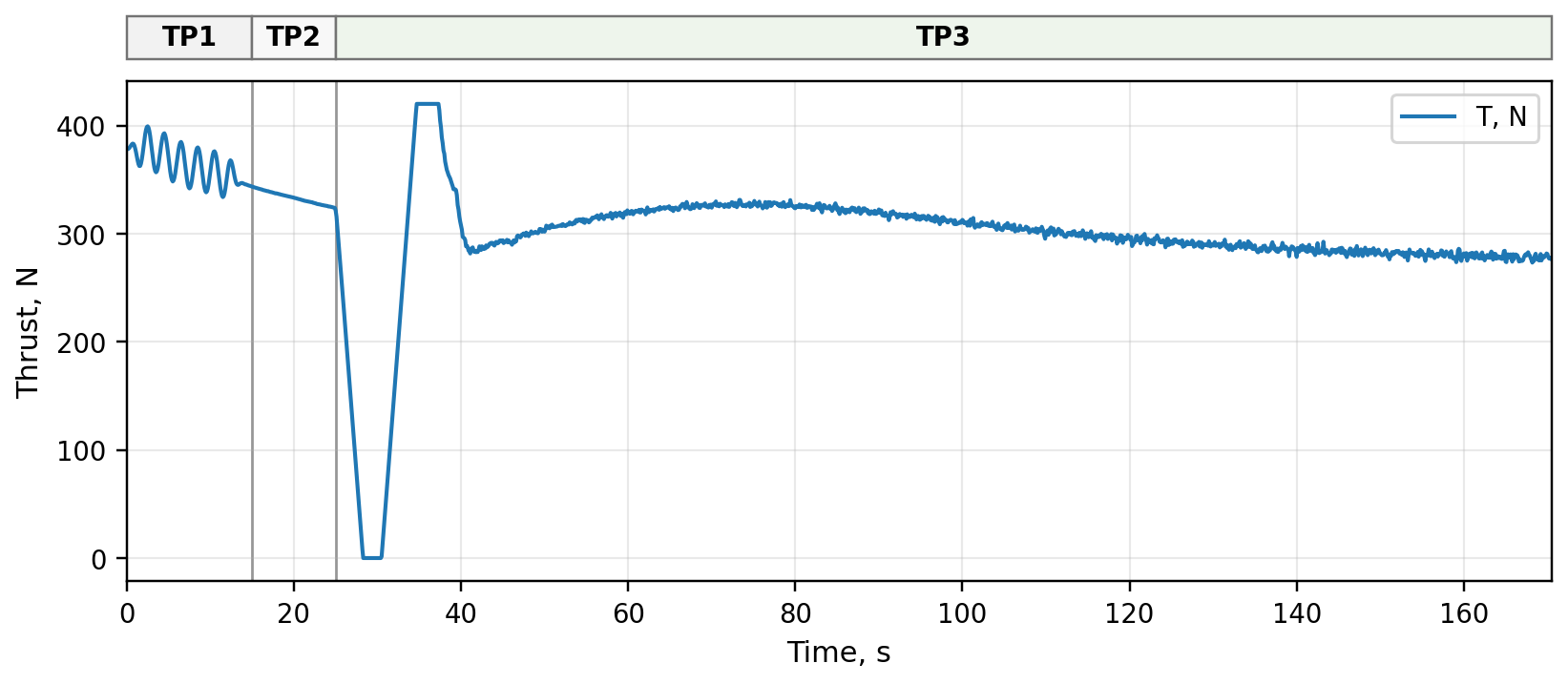}\\[4pt]
\includegraphics[width=0.90\textwidth]{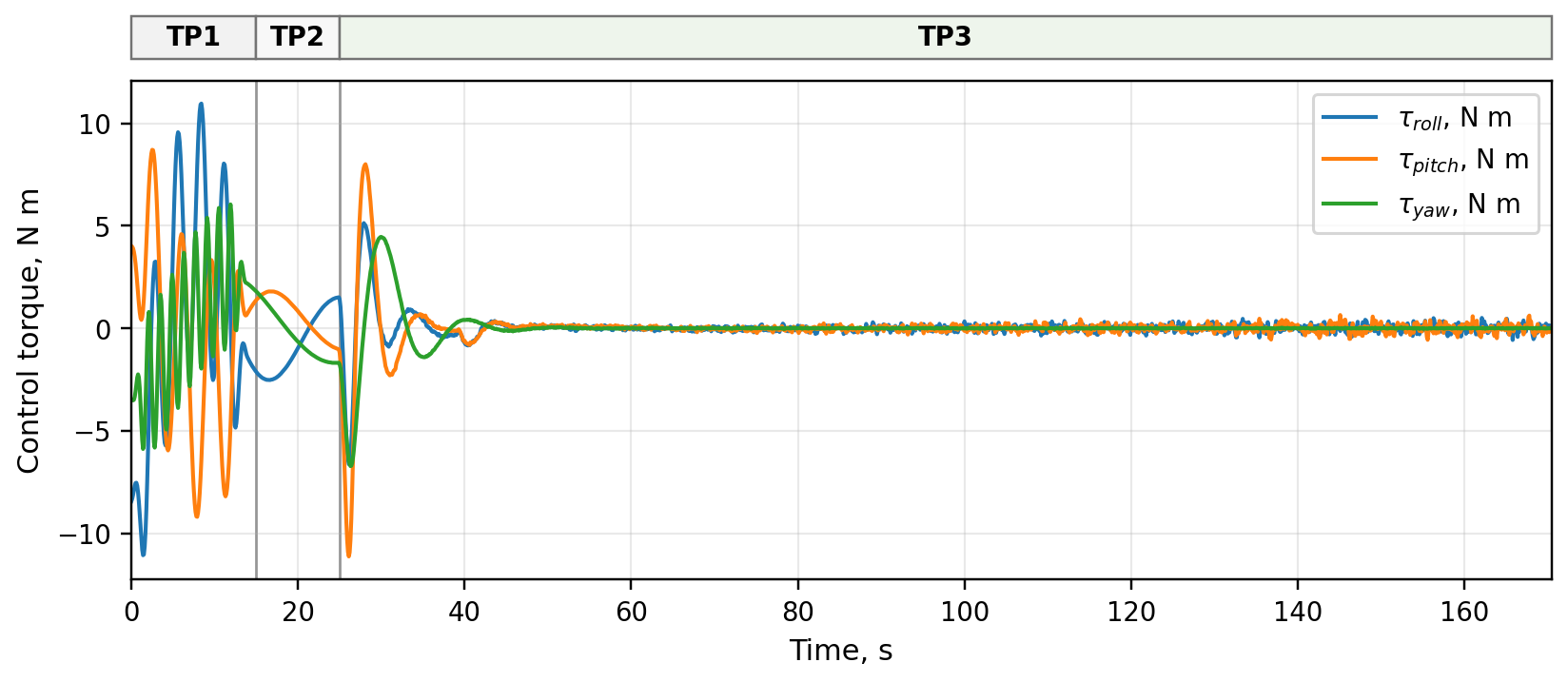}
\caption{All lunar-lander control inputs: main thrust $T$ and attitude-control torques $\tau_{\mathrm{roll}},\tau_{\mathrm{pitch}},\tau_{\mathrm{yaw}}$.}
\label{fig:lunar-inputs}
\end{figure}

The main thrust $T$ provides vertical braking and compensates lunar gravity. The torques $\tau_{\mathrm{roll}}$, $\tau_{\mathrm{pitch}}$, and $\tau_{\mathrm{yaw}}$ control attitude and therefore the inertial direction of the thrust vector. Their larger initial amplitudes correspond to rapid attitude correction; the pronounced restructuring near 25--40 s accompanies the transition to the main TP3 trajectory.

After approximately 50 s, the mean torque values are close to zero. A small residual high-frequency control ripple remains. Its amplitude is much smaller than the actuator bounds: after 100 s the roll and pitch torque standard deviations are approximately 0.17 and 0.18 N m, respectively, and their peak magnitudes remain below 0.65 N m versus the $\pm30$ N m limits. Such ripple can result from discrete feedback updates, measurement noise, active constraints, and compensation of small residual state or model errors; the plotted trajectories do not by themselves identify the relative contribution of these mechanisms. In this simulation it does not produce growing angular deviations. If additional smoothing is required, it can be reduced by a stronger $\Delta\bm{u}$ penalty, a suitable deadband or filter, or a smoother control parameterization.

\begin{figure}[H]
\centering
\includegraphics[width=0.90\textwidth]{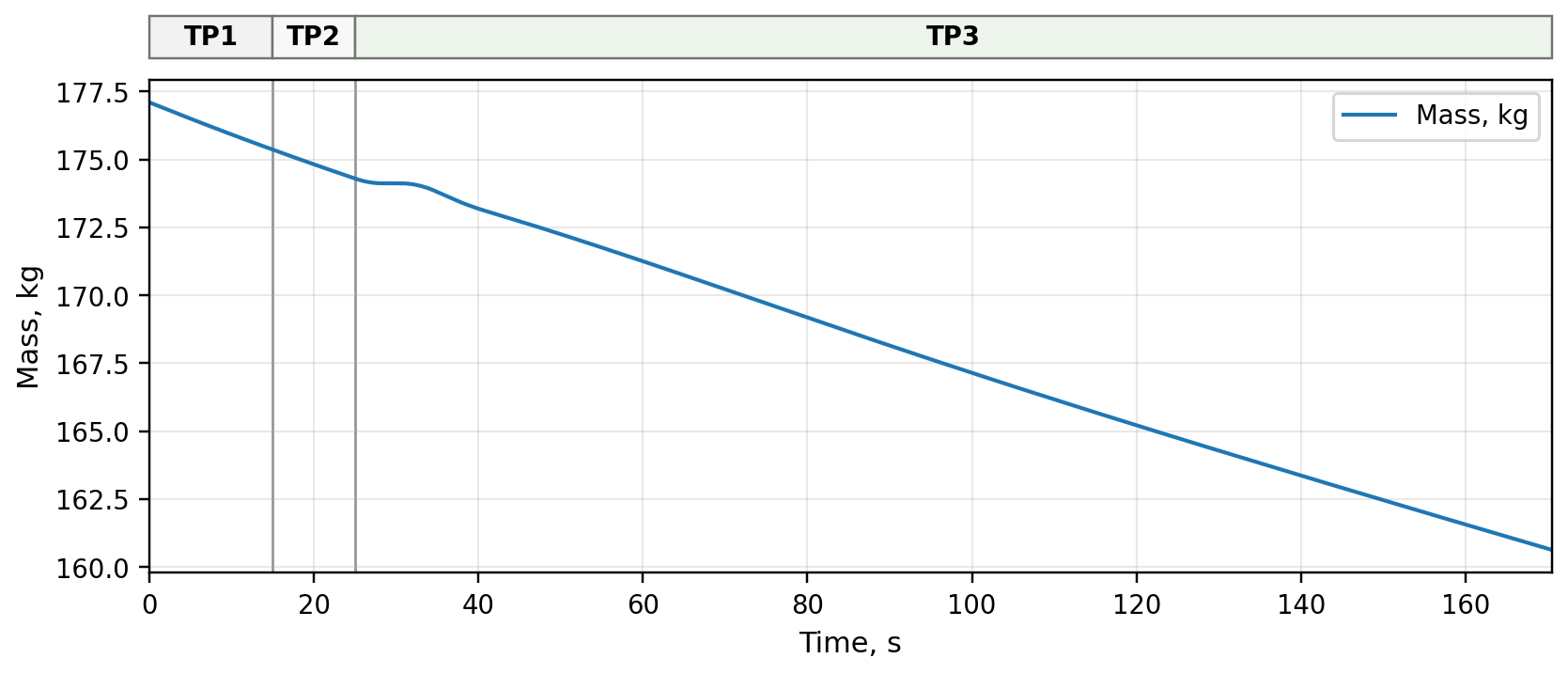}
\caption{Lunar-lander mass decrease due to propellant consumption.}
\label{fig:lunar-mass}
\end{figure}

The mass decreases monotonically from 177.1 kg to approximately 160.62 kg. Preparation of the global TP3 policy requires about 9.53 s from a 10 s computation budget. In this run, the global solve retains its initial valid policy because no improving iLQR step is accepted. The global policy is used for almost the entire TP3 interval. A local policy is successfully prepared at step 1690 and becomes active from step 1691, so the final 15 control intervals use the local policy while it is gradually blended with the preceding global policy.

\section{Extensions of the quadratic scheme}
AMIGO retains the quadratic dynamic-programming backbone described above and adds the mechanisms required for adaptive execution: state estimation, parameter identification, actuator limits, phase scheduling, and local replanning.

\begin{table}[H]
\centering
\caption{Main extensions of the baseline quadratic scheme.}
\small
\begin{tabularx}{\textwidth}{@{}lXX@{}}
\toprule
Aspect & Quadratic backbone & AMIGO \\
\midrule
Dynamics & linear model or sequential linearization & nonlinear RK4 transition and its Jacobians \\
Constraints & local quadratic problem & box constraints, active set, and rate limits \\
Measurements & state assumed available & continuous KF/EKF state estimation \\
Parameters & fixed model parameters & bounded LM identification with diagnostics \\
Execution & optimization over a given horizon & TP1--TP2--TP3, supervisory adaptive loop, and local iLQR/MPC \\
Solve time & usually outside the mathematical statement & explicit computation budget and use of the best available accepted policy \\
\bottomrule
\end{tabularx}
\end{table}

\section{Limitations and directions for development}
The present formulation has several general limitations that also suggest natural directions for further work.
\begin{enumerate}[leftmargin=*,itemsep=0.35em]
\item \textbf{Model uncertainty in the control objective.} Identification quality and prediction consistency are used mainly for supervisory decisions: whether an estimated model is sufficiently reliable for planning and control, whether TP1 should be repeated, or whether possible structural inadequacy should be declared. A more complete formulation could propagate parameter uncertainty and structural model mismatch directly into the optimization criterion, for example through robust, stochastic, or risk-sensitive extensions.
\item \textbf{Identifiability and excitation.} Reliable online identification requires that the measured trajectory contain enough information about the unknown parameters. Poor excitation, strongly correlated parameters, or short observation windows can lead to weakly determined estimates even when the numerical residual is small.
\item \textbf{Partial observation.} When only part of the state is measured, parameter identification can become coupled to the state estimator. In such cases it may be preferable to formulate identification residuals in measurement space, or to estimate states and parameters jointly.
\item \textbf{Local nature of iLQR.} The method is based on local linearization and a local quadratic approximation of the cost. For strongly nonlinear or highly nonconvex problems, convergence can depend on the initial nominal trajectory, regularization, and line-search strategy, and convergence to the global optimum is not guaranteed.
\item \textbf{General constraints.} Box constraints and control-rate limits are handled explicitly, but more general state, path, and coupled nonlinear constraints require additional machinery. Their treatment should remain compatible with the backward--forward structure and with real-time computation requirements.
\item \textbf{Finite computation time.} Receding-horizon optimization is performed under a finite computation budget. The achievable horizon length, update rate, and number of nonlinear iterations therefore depend on model complexity and available computing resources. A practical implementation should retain a valid previously accepted policy whenever a new optimization cannot be completed in time.
\end{enumerate}

\section{Conclusion}
AMIGO combines a quadratic LQR/iLQR optimizer with state estimation, online LM parameter identification, actuator constraints, and receding-horizon control. TP1 collects informative data and updates the model, TP2 prepares a long-horizon plan while the plant continues to move, and TP3 tracks the plan while periodically recomputing a local policy. A supervisory adaptive loop monitors predictive consistency during TP3. Persistent parametric mismatch can restart TP1 and TP2, whereas failure of re-identification to restore predictive consistency indicates possible structural model inadequacy and requires a predefined safe response rather than indefinite repetition of the same identification procedure.

The reverse-horizon quantities $P_{N-k}$ and $\bm{v}_{N-k}$ make the role of the backward pass explicit: every control law depends on the optimal cost of the future trajectory segment. The examples span a single identified parameter in the Van der Pol oscillator, several dynamic parameters in a quadcopter, and a variable-mass lunar-landing scenario with quaternion attitude and bounded thrust. A natural next step is to include model uncertainty directly in the control decision and to add explicit state and path constraints for safety-critical applications.

\section*{Acknowledgment}
ChatGPT (OpenAI) assisted with language editing and presentation of the manuscript. The author reviewed the technical content, calculations, and conclusions and remains responsible for them.


\begin{thebibliography}{9}
\footnotesize
\setlength{\itemsep}{0.25em}
\bibitem{ladnik2025}
I. Ladnik, ``A Quadratic Control Framework for Dynamic Systems,'' arXiv:2504.15396. \href{https://arxiv.org/abs/2504.15396}{arXiv:2504.15396}.

\bibitem{li2004}
W. Li and E. Todorov, ``Iterative Linear Quadratic Regulator Design for Nonlinear Biological Movement Systems,'' \emph{Proc. ICINCO}, pp. 222--229, 2004.

\bibitem{tassa2014}
Y. Tassa, N. Mansard, and E. Todorov, ``Control-Limited Differential Dynamic Programming,'' \emph{Proc. IEEE ICRA}, pp. 1168--1175, 2014.

\bibitem{bellman1957}
R. Bellman, \emph{Dynamic Programming}. Princeton University Press, 1957.

\bibitem{kalman1960}
R. E. Kalman, ``A New Approach to Linear Filtering and Prediction Problems,'' \emph{Journal of Basic Engineering}, 82(1), 35--45, 1960. DOI: 10.1115/1.3662552.

\bibitem{levenberg1944}
K. Levenberg, ``A Method for the Solution of Certain Non-Linear Problems in Least Squares,'' \emph{Quarterly of Applied Mathematics}, 2, 164--168, 1944. DOI: 10.1090/qam/10666.

\bibitem{marquardt1963}
D. W. Marquardt, ``An Algorithm for Least-Squares Estimation of Nonlinear Parameters,'' \emph{SIAM Journal on Applied Mathematics}, 11(2), 431--441, 1963. DOI: 10.1137/0111030.

\bibitem{nasaBeresheet}
NASA Science, ``Beresheet.'' \href{https://science.nasa.gov/mission/beresheet/}{NASA Science mission page}.

\bibitem{nammoLeros}
Nammo, ``LEROS 2b Apogee Engine.'' \href{https://www.nammo.com/products/space/space-propulsion-thrusters-and-engines/leros-2b/}{Nammo product page}.
\end{thebibliography}
\end{document}